\documentclass[twocolumn, tighten, times, twocolappendix]{aastex631}

\newcommand\hst{\textit{HST}}

\usepackage[caption=false]{subfig}
\usepackage{threeparttablex}
\usepackage{amsmath}
\usepackage{hyperref}
\hypersetup{linkcolor=blue,citecolor=blue,filecolor=blue,urlcolor=blue}

\shorttitle{Precise distances to the M31 system}
\shortauthors{A. Savino}
\graphicspath{{./}{figures/}}

\begin{document}


\title{The Hubble Space Telescope Survey of M31 Satellite Galaxies I. \\ RR Lyrae-based Distances and Refined 3D Geometric Structure}

\email{asavino@berkeley.edu}
\author[0000-0002-1445-4877]{Alessandro Savino}
\affiliation{Department of Astronomy, University of California, Berkeley, Berkeley, CA, 94720, USA}

\author[0000-0002-6442-6030]{Daniel R. Weisz}
\affiliation{Department of Astronomy, University of California, Berkeley, Berkeley, CA, 94720, USA}

\author[0000-0003-0605-8732]{Evan D. Skillman}
\affiliation{University of Minnesota, Minnesota Institute for Astrophysics, School of Physics and Astronomy, 116 Church Street, S.E., Minneapolis,
MN 55455, USA}

\author{Andrew Dolphin}
\affiliation{Raytheon, Tucson, AZ 85726, USA}

\author[0000-0002-3204-1742]{Nitya Kallivayalil}
\affiliation{Department of Astronomy, University of Virginia, 530 McCormick Road, Charlottesville, VA 22904, USA}

\author[0000-0003-0603-8942]{Andrew Wetzel}
\affiliation{Department of Physics and Astronomy, University of California, Davis, CA 95616, USA}

\author{Jay Anderson}
\affiliation{Space Telescope Science Institute, 3700 San Martin Drive, Baltimore, MD 21218, USA}

\author[0000-0003-0715-2173]{Gurtina Besla}
\affiliation{3 Department of Astronomy, University of Arizona, 933 North Cherry Avenue, Tucson, AZ 85721, USA}

\author[0000-0002-9604-343X]{Michael Boylan-Kolchin}
\affiliation{Department of Astronomy, The University of Texas at Austin, 2515 Speedway, Stop C1400, Austin, TX 78712, USA}


\author{James S. Bullock}
\affiliation{Department of Physics and Astronomy, University of California, Irvine, CA 92697 USA}

\author[0000-0003-0303-3855]{Andrew A. Cole}
\affiliation{School of Natural Sciences, University of Tasmania, Private Bag 37, Hobart, Tasmania 7001, Australia}

\author[0000-0002-1693-3265]{Michelle L.M. Collins}
\affiliation{Physics Department, University of Surrey, Guildford GU2 7XH, UK}

\author[0000-0003-1371-6019]{M. C. Cooper}
\affiliation{Department of Physics and Astronomy, University of California, Irvine, CA 92697 USA}

\author[0000-0001-6146-2645]{Alis J. Deason}
\affiliation{Institute for Computational Cosmology, Department of Physics, Durham University, Durham DH1 3LE, UK}

\author{Aaron L. Dotter}
\affiliation{Department of Physics and Astronomy, Dartmouth College, 6127 Wilder Laboratory, Hanover, NH 03755, USA}

\author{Mark Fardal}
\affiliation{Space Telescope Science Institute, 3700 San Martin Drive, Baltimore, MD 21218, USA}

\author{Annette M. N. Ferguson}
\affiliation{Institute for Astronomy, University of Edinburgh, Royal Observatory, Blackford Hill, Edinburgh, EH9 3HJ, UK}

\author[0000-0002-3122-300X]{Tobias K. Fritz}
\affiliation{ Department of Astronomy, University of Virginia, Charlottesville, 530 McCormick Road, VA 22904-4325, USA}

\author[0000-0002-7007-9725]{Marla C. Geha}
\affiliation{Department of Astronomy, Yale University, New Haven, CT 06520, USA}

\author[0000-0003-0394-8377]{Karoline M. Gilbert}

\affiliation{Space Telescope Science Institute, 3700 San Martin Drive, Baltimore, MD 21218, USA}
\affiliation{The William H. Miller III Department of Physics \& Astronomy, Bloomberg Center for Physics and Astronomy, Johns Hopkins University, 3400 N. Charles Street, Baltimore, MD 21218}

\author{Puragra Guhathakurta}
\affiliation{UCO/Lick Observatory, Department of Astronomy \& Astrophysics, University of California Santa Cruz, 1156 High Street, Santa Cruz, California 95064, USA}

\author{Rodrigo Ibata}
\affiliation{Observatoire de Strasbourg, 11, rue de l’Universite, F-67000 Strasbourg, France}

\author[0000-0002-2191-9038]{Michael J. Irwin}
\affiliation{Institute of Astronomy, University of Cambridge, Cambridge CB3 0HA, UK}

\author{Myoungwon Jeon}
\affiliation{School of Space Research, Kyung Hee University, 1732 Deogyeong-daero, Yongin-si, Gyeonggi-do 17104, Republic of Korea}

\author{Evan Kirby}
\affiliation{Department of Astronomy, California Institute of Technology, 1200 E California Boulevard, Pasadena, CA 91125, USA}
\affiliation{Department of Physics, University of Notre Dame, Notre Dame, IN 46556, USA}

\author[0000-0003-3081-9319]{Geraint F. Lewis}
\affiliation{Sydney Institute for Astronomy, School of Physics, A28,
The University of Sydney, NSW 2006, Australia}

\author[0000-0002-6529-8093]{Dougal Mackey}
\affiliation{Research School of Astronomy and Astrophysics, Australian National
University, Canberra 2611, ACT, Australia}

\author{Steven R. Majewski}
\affiliation{Department of Astronomy, University of Virginia, 530 McCormick Road, Charlottesville, VA 22904, USA}

\author{Nicolas Martin}
\affiliation{Observatoire de Strasbourg, 11, rue de l’Universite, F-67000 Strasbourg, France}
\affiliation{Max-Planck-Institut fur Astronomie, K\"{o}nigstuhl 17, D-69117 Heidelberg, Germany}

\author{Alan McConnachie}
\affiliation{NRC Herzberg Astronomy and Astrophysics, 5071 West Saanich Road, Victoria, BC V9E 2E7, Canada}
\affiliation{Physics \& Astronomy Department, University of Victoria, 3800 Finnerty Road, Victoria, BC V8P 5C2 Canada}

\author[0000-0002-9820-1219]{Ekta Patel}
\affiliation{Department of Astronomy, University of California, Berkeley, Berkeley, CA, 94720, USA}
\affiliation{Miller Institute for Basic Research in Science, 468 Donner Lab, Berkeley, CA 94720, USA}

\author[0000-0003-0427-8387]{R. Michael Rich}
\affiliation{Department of Physics and Astronomy, UCLA, 430 Portola Plaza, Box 951547, Los Angeles, CA 90095-1547, USA}

\author[0000-0002-4733-4994]{Joshua D. Simon}
\affiliation{Observatories of the Carnegie Institution for Science, 813 Santa Barbara Street, Pasadena, CA 91101, USA}

\author[0000-0001-8368-0221]{Sangmo Tony Sohn}
\affiliation{Space Telescope Science Institute, 3700 San Martin Drive, Baltimore, MD 21218, USA}

\author{Erik J. Tollerud}
\affiliation{Space Telescope Science Institute, 3700 San Martin Drive, Baltimore, MD 21218, USA}

\author[0000-0001-7827-7825]{Roeland P. van der Marel}
\affiliation{Space Telescope Science Institute, 3700 San Martin Drive, Baltimore, MD 21218, USA}
\affiliation{Center for Astrophysical Sciences,
  The William H. Miller III Department of Physics \& Astronomy,
  Johns Hopkins University, Baltimore, MD 21218, USA}



\begin{abstract}
We measure homogeneous distances to M31 and 38 associated stellar systems ($-$16.8$\le M_V \le$ $-$6.0), using time-series observations of RR Lyrae stars taken as part of the Hubble Space Telescope Treasury Survey of M31 Satellites. From $>700$ orbits of new/archival ACS imaging, we identify $>4700$ RR Lyrae stars and determine their periods and mean magnitudes to a typical precision of 0.01 days and 0.04 mag. Based on Period-Wesenheit-Metallicity relationships consistent with the \textit{Gaia}\ eDR3 distance scale, we uniformly measure heliocentric and M31-centric distances to a typical precision of $\sim20$~kpc (3\%) and $\sim10$~kpc (8\%), respectively. We revise the 3D structure of the M31 galactic ecosystem and: (i) confirm a highly anisotropic spatial distribution such that $\sim80\%$ of M31's satellites reside on the near side of M31; this feature is not easily explained by observational effects; (ii) affirm the thin (\textit{rms} $7-23$~kpc) planar ``arc” of satellites that comprises roughly half (15) of the galaxies within 300~kpc from M31; (iii) reassess physical proximity of notable associations such as the NGC 147/185 pair and M33/AND XXII; and (iv) illustrate challenges in tip-of-the-red-giant branch distances for galaxies with $M_V > -9.5$, which can be biased by up to 35\%. We emphasize the importance of RR Lyrae for accurate distances to faint galaxies that should be discovered by upcoming facilities (e.g., Rubin Observatory). We provide updated luminosities and sizes for our sample. Our distances will serve as the basis for future investigation of the star formation and orbital histories of the entire known M31 satellite system.

\end{abstract}



\section{Introduction} \label{sec:intro}

Satellite galaxies in the local Universe anchor our knowledge of low-mass galaxy formation and cosmology on small scales.  Their number counts, spatial distributions, 3D motions, chemical abundances, and star formation histories (SFHs) provide unique insight into a variety of physics including structure formation, cosmic reionization, and the nature of dark matter \citep[e.g.,][]{Hodge71, Rees86, Babul92, Mateo98, Moore99, Bullock00, Grebel03, Tolstoy09, Boylan-Kolchin11,Boylan-Kolchin12,Wetzel13,Brown14,Deason14,Wetzel16,Bullock17, Simon19}.

To date, most knowledge of low-mass galaxies comes from Milky Way (MW) satellites.  Their close proximities have enabled discovery and detailed characterization over a large dynamic range in stellar mass that is not possible to match in other environments \citep[e.g.,][]{Willman05,Kallivayalil06,Belokurov07,Simon07,Besla07,Kirby08,Kallivayalil13,VanderMarel14,Bechtol15,Koposov15,Fritz18,Simon19}.  Whilst substantial efforts are being made to identify and study low-mass satellites throughout the Local Volume in order to test the representative nature of the MW satellite population \citep[e.g.,][]{Chiboucas09, Dalcanton09,Calzetti15,Geha17,Smercina18,Crnojevic19,Bennet19,Okamoto19,Carlsten20, Drlica-Wagner21,Mao21,Carlsten22}, they are generally limited to fairly bright systems and coarse characterizations of their stellar populations \citep[e.g.,][]{DaCosta10,Weisz11,Cignoni19}. 

Our nearest large neighbor, the Andromeda galaxy (M31), occupies a special place in our quest to understand low-mass satellites. The M31 system is close enough that it is possible to study the morphology, stellar populations, abundances, dynamics, and gas properties of its constituent components in great detail.  But M31 is also quite different than the MW, providing a foil for comparison.  It is a more massive, metal-rich, evolved spiral galaxy \citep[e.g.,][]{Irwin05, Brown06, Kalirai06, Watkins10,Fardal13,Gilbert14,Mackey19}, with a different accretion history than the MW, as highlighted by its well-known prominent substructures \citep[e.g, ][]{McConnachie03, Zucker04,Martin06,Ibata07,Irwin08,Martin09,Richardson11,Martin13a,Martin13b,Martin13c,Bernard15,McConnachie18,Escala22}.


There have been a number of substantial efforts to provide high-quality data on the M31 satellites, in order to compare them to their MW counterparts.  Beyond mapping of the entire M31 area of the sky \citep[e.g., ][]{Ibata01,Ferguson02,Irwin05,Majewski07,McConnachie09}, significant investments with spectroscopic Keck instruments have produced resolved star metallicites, including $\alpha$-abundances in some cases, and velocity dispersions for most M31 satellites \citep[e.g.,][]{Geha06,Guhathakurta06,Geha10,Kalirai10,Tollerud12,Collins13,Vargas14,Gilbert19,Kirby20,Wojno20,Escala21}.  Recently, deep Hubble Space Telescope (\hst) imaging has produced high-fidelity SFHs for a small set of M31 satellites \citep[e.g.,][]{Geha15,Skillman17} and coarse SFHs for a much larger sample \citep[e.g.,][]{Weisz19b}. 

In \hst\ Cycle 27, our team was awarded 244 prime and 244 parallel \hst\ orbits to acquire deep imaging of 23 M31 satellites lacking color-magnitude diagrams (CMDs) that extended to the ancient main sequence turn-off (GO-15902, PI: D.\ Weisz).  These data are designed to fill in substantial gaps in our knowledge of the M31 satellite system, notably high-fidelity SFHs and precise distances for the entire population, while also providing first epoch images for proper motion measurements.  Combined with archival data, this program will help to characterize the M31 satellite population to a level of detail comparable to MW satellites.  

At the foundation of virtually all science related to M31 satellites are robust and uniformly measured distances.
 Many efforts have produced distances to individual galaxies in the M31 system \citep[e.g.,][and references therein]{McConnachie12}, while a handful of studies have published homogeneous distances for $\sim50$\% of the currently known M31 satellites \citep[e.g.,][]{McConnachie05, Conn12, Weisz19}. 

However, these distances come with limitations. Beyond the usual challenges of systematics introduced from small, heterogeneous analyses,  many of these measurements were based on the magnitude of the red giant branch tip (TRGB), either as a direct distance indicator or as an anchor to another distance proxy (such as the horizontal branch, HB, in \citealt{Weisz19}). However, the reliance on the TRGB limits the measurement robustness in low-mass systems which have sparsely populated RGBs \citep{Madore95}. Other complications include homogeneous and varied treatment of dust in the face of dust substructures around M31 \citep[e.g., ][]{Ruoyi20} and mitigating systematics such as transforming between ground and \hst-based filter systems \citep[e.g.,][]{Riess18,Riess21}.  These types of uncertainties, which can be of order 0.1~mag (i.e., 5\% of the measured distances) or more, can quickly become the dominant source of uncertainty, particularly in the limit of excellent data quality.

Many of these limitations can be overcome by computing distances based on RR Lyrae \citep[e.g., ][]{Liu90,Chaboyer99,Bono01}.  As old, low-mass stars, RR Lyrae appear to exist in virtually all known, resolved galaxies in which they can be detected \citep[e.g.,][and references therein]{Catelan04,Bernard09,Clementini14,Martinez-Vazquez19}, meaning they can provide distances for galaxies of any luminosity.  They also have the advantage of well-established reddening-independent scaling relations that can be used to infer robust distances \citep{Madore82} and more recently, have been anchored to the geometric distance scale of \textit{Gaia} parallaxes \citep[e.g.,][]{Neeley19,Nagarajan22,Garofalo22}.  

However, RR Lyrae are challenging to observe, particularly at large distances.  They can be quite faint ($M_V \sim +0.5$) and they require highly cadenced observations for robust period and mean magnitude determinations. Within the M31 ecosystem, these observational demands require the heavily over-subscribed capabilities of \hst, meaning that to date, only $\sim$30\% of galaxies affiliated with M31 have RR Lyrae distances \citep[e.g.,][]{Pritzl02,Brown04,Pritzl04,Pritzl05, Yang10,Jeffery11,Yang12,Cusano13,Cusano15,Cusano16,Martinez-Vazquez17}.



Our \hst\ survey of M31 satellites was designed to provide the cadenced data needed for RR Lyrae-based distances to all galaxies. In this paper, we use uniformly reduced new and archival \hst\ imaging to derive homogeneous RR Lyrae-based distances to virtually all known dwarf galaxies presently (or that might have been in the past) within the M31 virial radius, including a new RR Lyrae distance to M31 itself, which serves to anchor the 3D geometry of the M31 system. 


This paper is organized as follows. We describe the data used in this paper in \S~\ref{sec:data}.  We present the variable star analysis in \S~\ref{sec:variables} and the distance determinations in \S~\ref{sec:distances}. We validate our results in \S~\ref{sec:sys}, and in \S~\ref{sec:applications} we use our distances to refine the geometry of the M31 satellite system and explore a handful of science cases that are sensitive to new distances. 
\linebreak

\section{Data} \label{sec:data}
This paper makes use of 244 orbits of \hst\ imaging acquired between Oct. 2019 and Oct. 2021 as part of the \hst\ Treasury Survey of the M31 satellite galaxies (GO-15902; PI:\ D. Weisz).  We will present a full survey overview in a forthcoming publication (Weisz et al., in prep.); here we summarize the data components relevant to the RR Lyrae. Our data consist of deep ACS/WFC F606W and F814W imaging of 23 dwarf galaxies within approximately 300~kpc from M31. We designed observations to reach the oldest main sequence turn-off of our targets and, in doing so, ensure a cadence that enables short-variability analysis.  We obtained parallel F606W and F814W WFC3/UVIS imaging.  However, for the purposes of RR Lyrae distances, we only focus on the ACS fields, which are generally more populated.

The program was designed to get at least 10 \hst~orbits per galaxy.  This minimum orbit allocation provided sufficient cadence  needed for RR Lyrae distance determinations and coarse separation into types (e.g., RRab vs. RRc) while also ensuring our depth goals were met. However, due to updated distances that were finalized after proposal acceptance and the need for some galaxies to be scheduled during times of higher than anticipated background (e.g., to ensure timely program completion), the Telescope Time Review Board (TTRB) allowed us to reallocate orbits between a handful of galaxies to ensure adequate depth in the most affected systems.  As a result, a handful of galaxies (And~{\sc X}, And~{\sc XXIV}, and And~{\sc XXX}) were observed with less than 10 orbits.  Though sub-optimal for the RR Lyrae-based science in the proposal, the lower cadence was still adequate for reasonable distance determinations. However, for several orbits, \hst\ failed to acquire guide stars, rendering the data unusable.  As we will describe in the main survey paper, the TTRB usually granted replacement of lost orbits. The exceptions to this were And~{\sc XIII}, And~{XX}, and And~{\sc XXIV}, which have a lower than anticipated number of epochs.


We complemented our new data with a compilation of archival ACS/WFC observations that have targeted the M31 system over the years. For much of our sample, we are able to include single orbit ACS/WFC F606W and F814W observations (GO-13699, PI: N. Martin) that spatially overlap our new observations.  Most importantly, this adds additional epochs to the RR Lyrae light curves. Our program did not target M31 satellites that already have existing ACS/WFC imaging of similar (or better) depth and cadence. For these galaxies, we uniformly reduced and analyzed F475W, F606W, and F814W imaging from GO-9392 (PI:\ M. Mateo), GO-10505 (LCID, PI:\  C. Gallart), GO-10794/11724 (PI:\ M. Geha), GO-13028/13739 (ISLAndS, PI:\ E. Skillman), GO-13738 (PI:\ E. Shaya), GO-13768 (PI:\ D. Weisz), GO-14769/15658(PI: \ S. Sohn) and GO-15302 (PI:\ M. Collins).

Finally, we included \hst\ imaging of the two large galaxies in the M31 system: M33 and M31 itself. For M33 we used data from program GO-10190 (PI:\ D. Garnett). Specifically we used the closest field to the center of M33. We chose this field as, among the existing deep ACS imaging of M33, it is the one with the most extensive time-series coverage while also having the highest number of known RR Lyrae \citep{Yang10,Tanakul17}. For M31, we used 114 orbits from program GO-9453 (PI:\ T. Brown). This field, which is placed in the inner halo of M31, is excellent to serve as an anchor for the M31 distance as it is relatively close to M31's center (11~kpc, projected) provides exquisite time-series sampling and hosts a substantial population of RR Lyrae stars \citep{Jeffery11}. We also included a field placed on M31's most notable tidal feature, the Giant Stellar Stream (GSS), using data from program GO-10265 (PI:\ T. Brown).

\begin{table*}
\centering
\caption{A list of our targets, total \hst\ exposure time, number of photometric epochs, and ID of the original observing programs. Roughly 60\% of the galaxies and 35\% of the total exposure time in this dataset were newly acquired as part of GO-15902.}
\begin{tabular}{llrcccc}


\toprule
Galaxy ID &Other Names&$t_{exp}$ [s]&       F475W Epochs  & F606W Epochs & F814W Epochs & \hst\ Proposal ID \\
\toprule

         M31&Andromeda, NGC 224, UGC 454&141 305&...&58&60&9453\\
         GSS&&133 160&...&44&64&10265\\

         M32&NGC221, UGC 452&59 036&...&16&28&9392,15658\\

         M33&Triangulum, NGC 598, UGC 1117&49 780&...&20&24&10190\\

         NGC147&DDO 3, UGC 326&98 464&...&44&36&10794,11724,14769\\

         NGC185&UGC 396&78 782&...&36&30&10794,11724,14769\\

         NGC205&M110, UGC 426&21 957&...&11&11&15902\\

         And~{\sc I}&&51 964&22&...&22&13739\\

         And~{\sc II}&&40 268&17&...&17&13028\\

         And~{\sc III}&&51 964&22&...&22&13739\\

         And~{\sc V}&&21 988&...&11&11&15902\\

         And~{\sc VI}&Peg dSph&22 070&...&11&11&15902\\

         And~{\sc VII}&Cas dSph&24 425&...&12&12&15902\\

         And~{\sc IX}&&24 362&...&13&13&13699,15902\\

         And~{\sc X}&&19 907&...&11&11&13699,15902\\

         And~{\sc XI}&&22 065&...&11&11&15902\\

         And~{\sc XII}&&21 896&...&11&11&15902\\

         And~{\sc XIII}&&17 595&...&9&9&15902\\

         And~{\sc XIV}&&22 065&...&11&11&15902\\

         And~{\sc XV}&&40 216&17&...&17&13739\\

         And~{\sc XVI}&&30 816&13&...&13&13028\\

         And~{\sc XVII}&&37 564&...&20&20&13699,15902\\

         And~{\sc XIX}&&40 504&...&14&18&15302\\

         And~{\sc XX}&&20 096&...&11&11&13699,15902\\

         And~{\sc XXI}&&28 776&...&15&15&13699,15902\\

         And~{\sc XXII}&&35 454&...&18&18&13699,15902\\

         And~{\sc XXIII}&&24 321&...&13&13&13699,15902\\

         And~{\sc XXIV}&&17 874&...&10&10&13699,15902\\

         And~{\sc XXV}&&26 746&...&14&14&13699,15902\\

         And~{\sc XXVI}&&31 096&...&16&16&13699,15902\\

         And~{\sc XXVIII}&&47 240&20&...&20&13739\\

         And~{\sc XXIX}&&26 184&...&14&14&13699,15902\\

         And~{\sc XXX}&Cas~{\sc II}&19 976&...&11&11&13699,15902\\

         And~{\sc XXXI}&Lac~{\sc I}&28 834&...&15&15&13699,15902\\

         And~{\sc XXXII}&Cas~{\sc III}&33 400&...&16&16&15902\\

         And~{\sc XXXIII}&Per~{\sc I}&24 364&...&13&13&13699,15902\\

         Pisces&LGS 3&58 416&12&...&36&10505,13738\\

         Pegasus DIG&DDO 216, UGC 12613&71 670&29&...&29&13768\\
         
         IC1613&DDO 8, UGC 668&58 608&24&...&24&10505\\
         \toprule

\end{tabular}

\label{tab:log}
\end{table*}

Table~\ref{tab:log} lists the 39 stellar systems in our sample, which includes virtually all known galaxies within the virial radius of M31 \citep[we adopt 266~kpc, from][]{Fardal13,Putman21} and several other galaxies at larger radii. The only notable  exceptions within the virial radius of M31 are IC10, for which even the deepest available \hst\ observations \citep[GO-10242]{Cole10} are not able to pierce through the thick foreground dust extinction and allow RR Lyrae identification, and Peg~{\sc V} \citep{Collins22} which has been recently discovered and therefore still lacks deep \hst\ imaging. Other notable stellar systems that are not included in this paper are And~{\sc XVIII}, whose distance of $1.33^{+0.06}_{-0.09}$~Mpc \citep{Makarova17} places it far beyond the virial radius of M31, and And~{\sc XXVII}, which is currently thought to be a tidally disrupted structure \citep{Collins13}.

We measured stellar fluxes for individual stars by using the point spread function (PSF) crowded field photometry package  \texttt{DOLPHOT} \citep{Dolphin00,Dolphin16}, which has been widely used for \hst\ resolved star studies throughout the Local Group and Local Volume \citep[e.g., ][]{Dalcanton09,Dalcanton12,Gallart15,Martin17,Williams21}.  We adopted the same \texttt{DOLPHOT} setup developed for the Panchromatic Hubble Andromeda Treasury program and detailed in \citet{Williams14}, as it is the optimal configuration for the typical crowding level of our images. Our forthcoming survey paper will provide extensive details on data reduction and photometry validation tests. Making use of \texttt{DOLPHOT}, we analyzed the individual ACS exposures, performing simultaneous PSF-photometry on all \texttt{flc} frames. This process results in a deep photometric catalog for each galaxy with time-tagged stellar flux measurements for each epoch. We used these catalogs as a basis for our variability analysis and will include them in our public data release alongside the survey paper.

\begin{table}
\centering
\caption{Parameters and priors used in the light-curve modeling. Note that $\Phi$, $A$, and $M$ are parameters for each filter.}
\begin{tabular}{lll}


\toprule
Parameter&Prior&Description\\   
\toprule
P [d]&$\mathcal{U}(0,1.2)$&Period [d]\\
$\Phi_{\lambda}$&$\mathcal{U}(0,1)$&Pulsation phase\\
$A_{\lambda}$&$\mathcal{U}(0,1.5)$&Pulsation amplitude [mag]\\
$M_{\lambda}$&$\mathcal{U}(23,26)$&Intensity-averaged magnitude\\

\toprule

\end{tabular}

\label{tab:RRprior}
\end{table}

\pagebreak
\section{Variable identification and modeling} \label{sec:variables}

\subsection{Methodology}
\label{sec:RRmod}

In this section we summarize the main steps of our RR Lyrae detection and modeling algorithm and describe the broad properties of the RR Lyrae sample. We will present an in-depth description of the variable star sample and analysis (e.g., variable star completeness quantification, non-RR Lyrae variables) in the main survey paper. Here, we provide a description of our procedure as applied to putative RR Lyrae. 

\begin{figure}

     \subfloat
	{\includegraphics[width=0.45\textwidth]{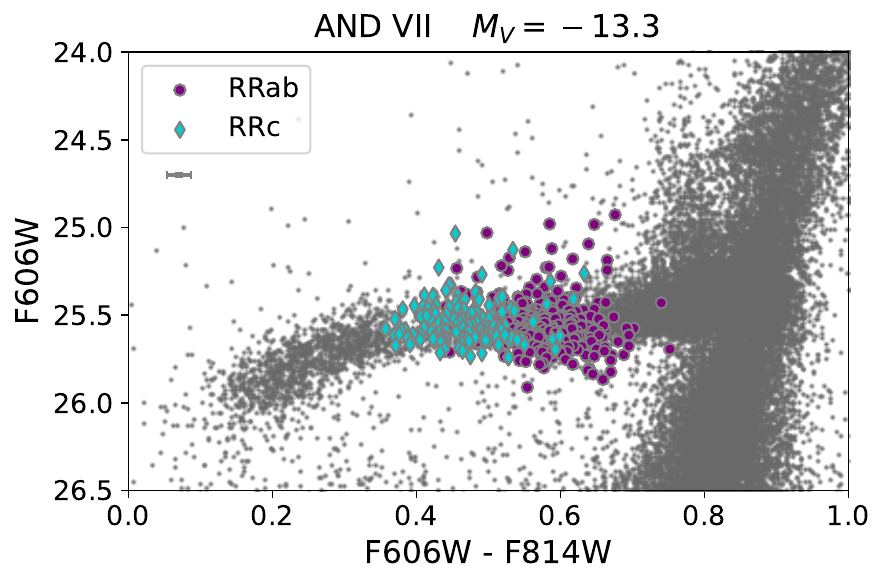} \label{fig:A7CMD}} \quad
	 \subfloat
	{\includegraphics[width=0.45\textwidth]{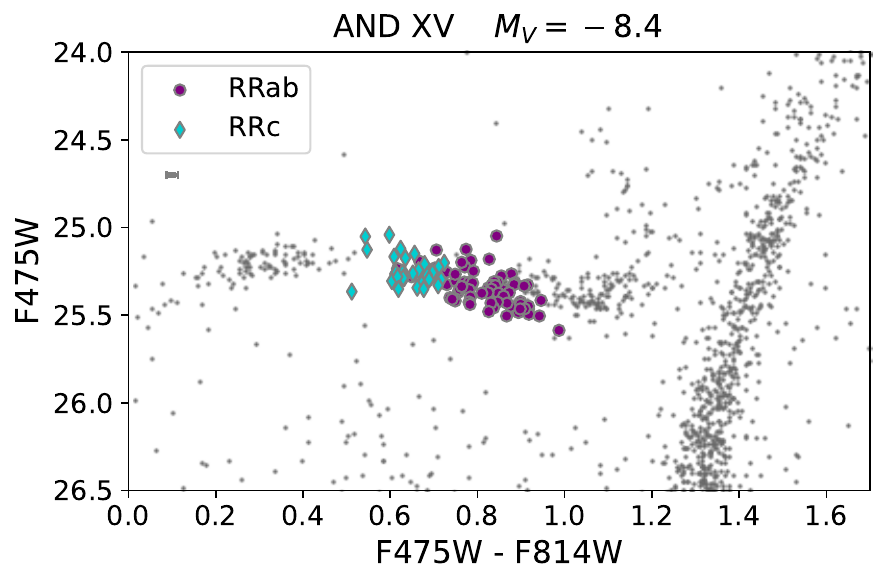} \label{fig:A15CMD}} \quad
	\subfloat
	{\includegraphics[width=0.45\textwidth]{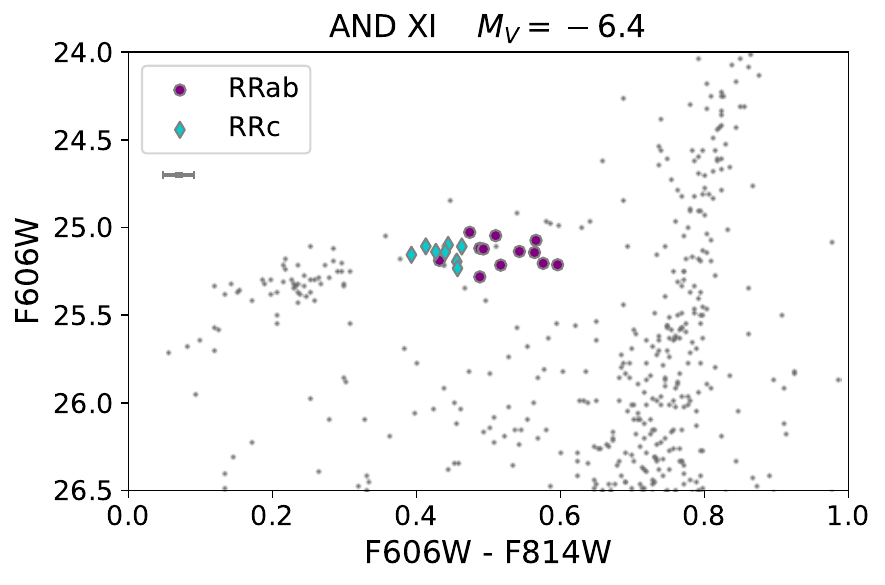} \label{fig:A11CMD}} \quad
\caption{Example CMDs for three targets (And~{\sc VII}, And~{\sc XV}, And~{\sc XI}) of our sample, zoomed in on the HB. These examples illustrate the range of galaxy stellar mass, RR Lyrae population, and data quality across our sample. Confirmed RR Lyrae stars are represented as purple circles (RRab) and cyan diamonds (RRc). The error bars in the upper right corner show the median photometric uncertainties at the level of the HB. }
\label{fig:CMDs}
\end{figure}

We developed an approach to address two main challenges provided by our dataset: the high number of RR Lyrae candidates (several thousands) and the modest number of photometric epochs that many of our target galaxies have (between 10 and 15 per filter, for roughly half of our sample; see Table~\ref{tab:log}). This required the development of a procedure that could be carried out mostly in an automated fashion, while being robust enough to reliably model light curves in a sparsely sampled regime. We devised the following multi-step approach:
\begin{itemize}
    \item [i)]{We first identified all possible variable stars candidates following the procedure described in \citet{Dolphin01,Dolphin04}, which selects variable candidates on the basis of four criteria, namely: (1) the $rms$ of the magnitude measurements, (2) the amount of crowding contamination reported by \texttt{DOLPHOT}, (3) the $\chi^2$ of the PSF fits, and (4) a variability metric based on \citet{Lafler65}.  This step in the procedure is commonly used in the literature \citep[e.g., ][for further details]{McQuinn15}.}

    \item [ii)]{We then inspected the variable sources on the CMD, where we selected likely RR Lyrae candidates, i.e., all variable stars lying within roughly half a magnitude from the mean magnitude of the HB and within the approximate color range of the instability strip, i.e. $0.45\lesssim {(F475W-F814W)}\lesssim1.1$ and $0.15\lesssim {(F606W-F814W)}\lesssim0.7$, adjusted for extinction \citep{Schlafly11}.}  
    \item [iii)]{For each star, we made a first estimate of the pulsation period using the peak of the variability merit function defined by \citet{Saha17}. This method is optimized for the analysis of sparse multi-band light curves and combines metrics from the two main families of periodicity analysis, Fourier-based periodograms and phase-dispersion minimization, to boost the pulsation signal relative to the artifacts introduced by the discrete data sampling.}
    \item [iv)]{We used the period measured in the previous step as a first guess for a multi-band fitting of the \hst\ light curve using 57 empirical templates derived from the extensive ground-based observations presented in \citet{Monson17}. Our light-curve models are parametrized by the pulsation period $P$, as well as a set of filter-dependent pulsation phases ($\Phi_{F606W/F475W}$ and $\Phi_{F814W}$), amplitudes ($A_{F606W/F475W}$ and $A_{F814W}$) and intensity-averaged magnitudes ($M_{F606W/F475W}$ and $M_{F814W}$). We performed the light-curve fit in the native \hst\ bands. We constructed our models using ground-based templates in the most closely matching filter to our \hst\ observations, i.e., Johnson B for F475W, Johnson V for F606W, and Johnson I for F814W. The effective wavelengths of the ground-based filters are sufficiently similar to the \hst\ counterparts that they result in similar light-curve shapes, while most of the difference between ground-based and \hst\ light curves is effectively bypassed by our free parameters (amplitudes and mean magnitudes). Using a Gaussian likelihood function and non-informative priors (see Table~\ref{tab:RRprior}) we sampled the posterior probability distribution (PPD) for each star using the affine invariant ensemble Markov chain Monte Carlo (MCMC) sampler \texttt{emcee} \citep{Foreman-Mackey13}. We defined the convergence length of the MCMC chain as 50 times the auto-correlation length. This sampling provides constraints on each star's variability parameters (defined as the 50th percentile of the PPD), the pulsation mode (RRab or RRc, based on the period and the best-fitting template), and a full uncertainty characterization of our measurements (15.9th and 84.1th percentiles of the PPD).}
    
    \item [v)]{We inspected the output sample using several diagnostic diagrams such as the period-amplitude diagram and the amplitude-amplitude diagram. We flagged $\sim10$\% of the sample as anomalous sources in at least one of these diagrams. We followed up with manual inspection and refinement of the fit, when possible. We discarded sources that show obvious modeling inconsistencies or ambiguous period solutions (e.g., significant multimodality in the variability function of step iii).  For the purpose of distance determinations, we prioritized purity over completeness, but for forthcoming population studies we will re-visit completeness (e.g., using artificial variable stars). As our light-curve sampling is not always sufficient to recover secondary pulsation modes, we did not attempt to identify RRd variables, which were either assigned to the RRab/RRc sample (based on the dominant pulsation mode) or discarded due to the poor light-curve fitting.  We discuss this in more depth in the main survey paper.}
\end{itemize}
Step (i) of this procedure produced a catalog of 6190 candidate variable sources.  Application of the subsequent steps (ii to v) refined this to 4775 \textit{bona fide} RR Lyrae, which we use for the distance determinations. We detect RR Lyrae in each system analyzed, ranging from a minimum of 4 (And~{\sc XIX}) to a maximum of 712 (Peg~DIG).


\begin{figure}

     \subfloat
	{\includegraphics[width=0.45\textwidth]{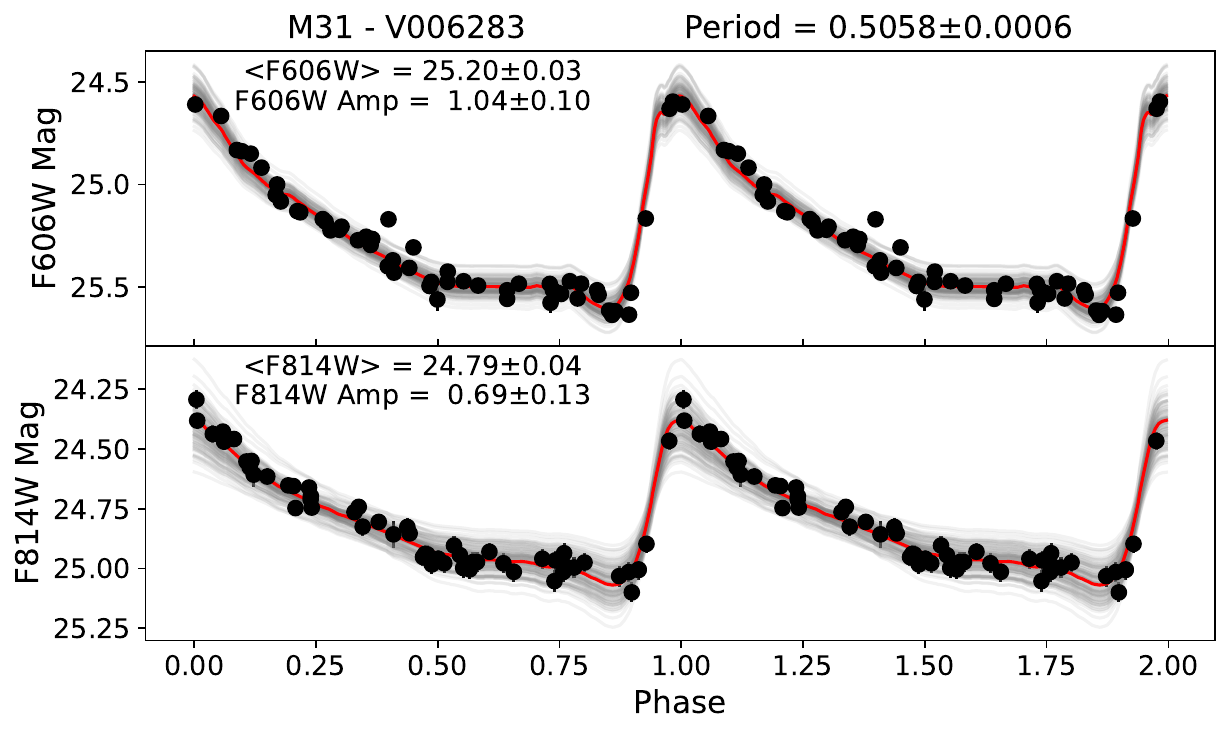} \label{fig:LCdense}} \quad
	 \subfloat
	{\includegraphics[width=0.45\textwidth]{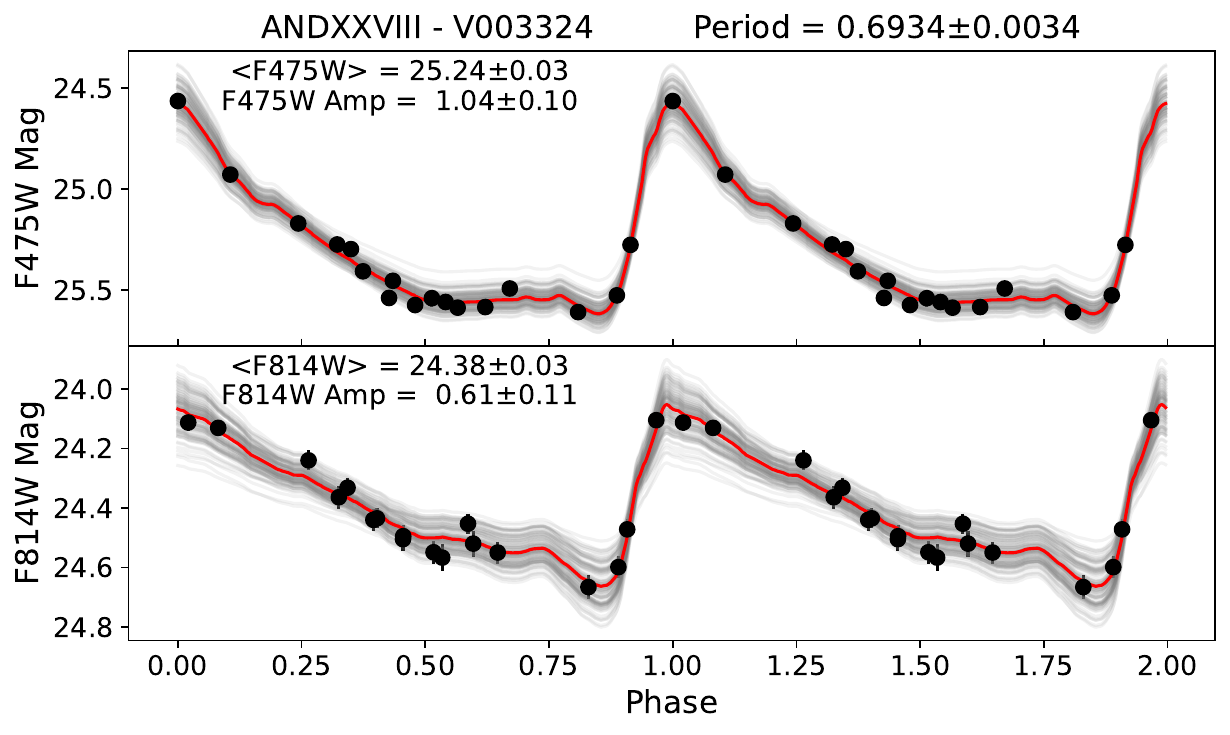} \label{fig:LCmed}} \quad
	\subfloat
	{\includegraphics[width=0.45\textwidth]{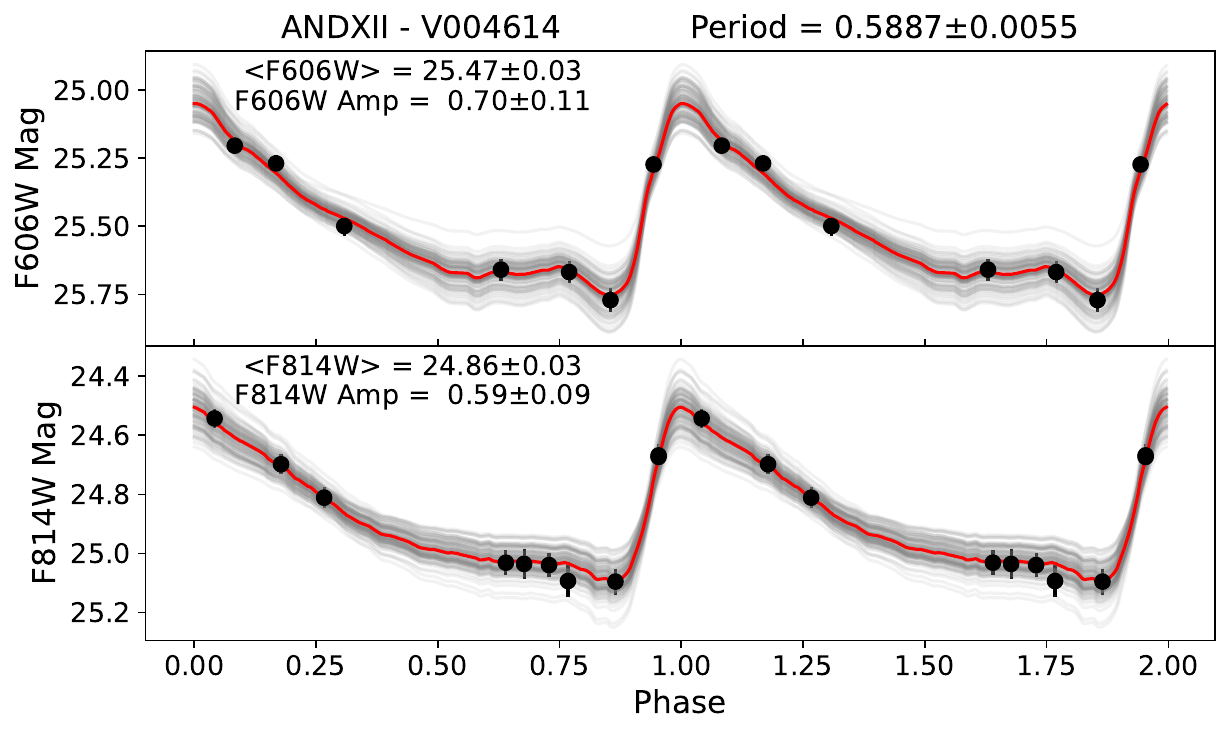} \label{fig:LCsparse}} \quad
\caption{Observed light curves (black points) of three RR Lyrae stars within our sample of 4694 variables, demonstrating representative time-series samplings for our dataset. The best fit template (red line) is overlaid on the observed light curves. The grey lines show 100 random samplings from our MCMC chains, providing an illustration of our model uncertainties.  The models are well-matched to the data in all cases. The titles and insets list the constrains on the period, amplitude, and mean magnitude of the RR Lyrae star, marginalized over all other model parameters.}
\label{fig:LC}
\end{figure}

\begin{figure*}

\plotone{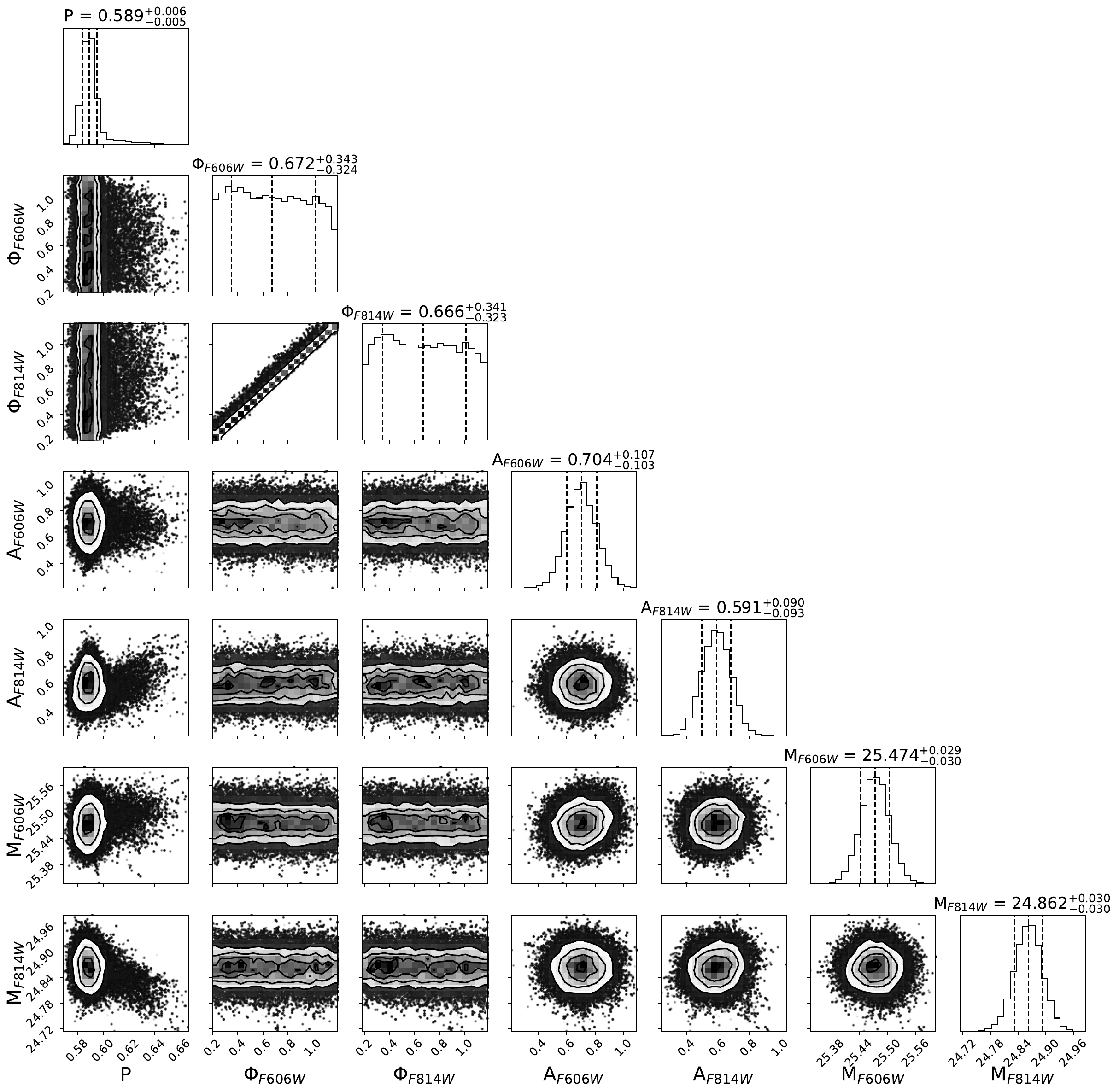}
\caption{Corner plot illustrating the posterior distribution for our variability model of RR Lyrae And~{\sc XII} - V004614, one of the sparsest sampled light curves in our dataset. In addition to the period (P), the light curve is parametrized through a phase ($\Phi$), an amplitude ($A$) and an intensity-averaged magnitude ($M$), in each filter. The vertical dashed lines and the plot headers report the 50th, 15.9th, and 84.1th percentiles of the PPD. This is a prototypical example of our posteriors, illustrating that variability parameters are generally well-constrained (except for the phase parameters which we consider nuisance parameters to be marginalized over) even for our more sparsely sampled data.}
\label{fig:MCMC}
\end{figure*}

\subsection{RR Lyrae Sample and Characteristics}

 Figure~\ref{fig:CMDs} shows CMDs for three galaxies (And~{\sc VII}, And~{\sc XV} and And~{\sc XI}), and the RR Lyrae stars we identified in them.  These examples illustrate the data quality and RR Lyrae population size across the full galaxy luminosity range of our sample. The CMDs highlight the excellent photometric quality of our data. At the magnitude of the HB ($m_{F606W} \sim 25$), the typical photometric error for non-variable sources is 0.01~mag, for single epoch variables 0.03~mag, and the photometric completeness within our observed fields is virtually 100\%.
 
 Figure~\ref{fig:LC} shows a range of example light curves and best fit models, selected to illustrate the range of data and fit quality.  Even in the limit of our sparsest sampling (e.g., And~{\sc XII}), our light curve fits are generally good. We constrain periods to a typical precision of 0.01~days, pulsation amplitudes to within 0.11~mag, and the mean intensity averaged magnitudes to 0.04~mag. 
 
 Figure~\ref{fig:MCMC} shows that we generally have robust fits even in the sparsely sampled regime.  Here, we show the PPDs for our fit to And~{\sc XII} - V004614 (bottom light curve in Fig.~\ref{fig:LC}), a typical example of our lowest sampling regime. The posteriors are generally well-constrained and unimodal, except for the phase $\Phi$. The purpose of this phase term is to ensure that, for any given period, the maximum-light epochs of model and data are matching. Therefore, $\Phi$ is effectively a nuisance parameter over which we marginalize. In our lowest light-curve sampling regime, multi-modality in the other variability parameters occasionally occurs due to trade-offs between light curve cadence and variable star period.  Most of these stars were flagged as anomalous in our step (v) and re-fit or removed.  As a guard against poorly modeled stars that made it past this step, we used a Gaussian Mixture Model (GMM, see \S \ref{sec:Model}) to mitigate the impact of contaminants on our distance determinations.
 
\begin{figure*}
\plotone{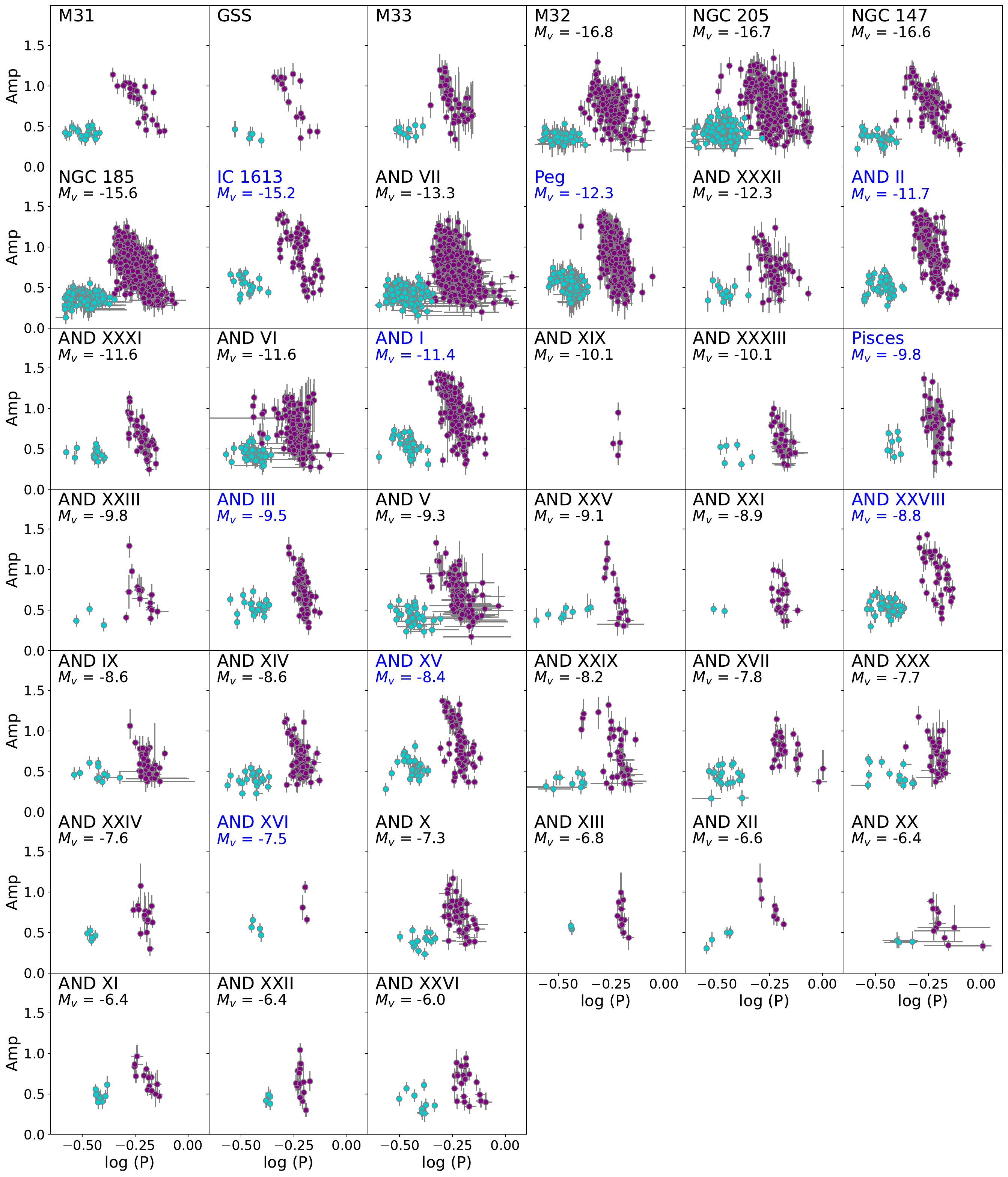}

\caption{RR Lyrae period-amplitude diagrams for our entire sample, ordered by decreasing galaxy luminosity.  Galaxy labels are color-coded by whether the amplitude refers to the F475W (blue) or F606W (black) magnitude variation. The purple and cyan points indicate RRab (fundamental mode) and RRc (first-overtone). This homogeneous compilation of period-amplitude diagrams illustrates the diversity of RR Lyrae populations in our galaxy sample, which, in turn, is indicative of a variety of formation histories.}
\label{fig:Bailey}
\end{figure*}

Figure~\ref{fig:Bailey} shows the period-amplitude diagrams \citep[Bailey diagrams;][]{Bailey19} for our sample. This diagram is a widely used diagnostic of RR Lyrae properties \citep[e.g.,][]{Sandage81,Soszynski09,Clementini19} and our ability to recover well-defined fundamental-mode (RRab, shown in purple) and first-overtone (RRc, in cyan) sequences illustrate the reliability of our light-curve modeling scheme. 

This gallery of Bailey diagrams is the largest homogeneously derived of its type and the first for a nearly complete satellite system.  It shows the sheer diversity of the RR Lyrae populations, both in size, RRab/RRc ratio and morphology of the period-amplitude sequences. In terms of RR Lyrae population size, there is a general trend between decreasing galaxy luminosity and decreasing number of RR Lyrae stars. However, at fixed galaxy luminosity, there is substantial scatter in the number of  RR Lyrae and several factors may be contributing. 

As RR Lyrae are primarily the manifestation of old stars that live in a narrow temperature range on the HB \citep[e.g.,][]{Walker89,Savino20}, their number abundance and pulsation properties are strongly influenced by the SFH of their host galaxy, which in turn affects the morphology of the HB \citep[e.g., ][]{Salaris13,Savino18,Savino19}. A clear demonstration of this effect can be appreciated, for instance, by comparing NGC 147 and NGC 185. While NGC 147 is about a magnitude brighter than NGC 185, we detect roughly five times fewer RR Lyrae (124 versus 579) in the former than in the latter. This is explained by NGC 147 having a substantially younger stellar population than NGC 185 \citep{Geha15}, and therefore a much lower RR Lyrae specific frequency (number of RR Lyrae produced per unit luminosity). Similar SFH effects also have consequences on the RRab/RRc ratio, since efficient production of RRc pulsators requires a sufficient extension of the HB to high temperatures. On the other hand, the presence of a predominantly red HB \citep[as in the case of some of our targets,][]{Martin17} will result in a deficiency of first overtone pulsators.

Another feature we note is the degree of variation in the slope of the RRab period-amplitude sequence, which can range from substantially slanted (e.g., NGC 147) to almost vertical (e.g., And~{\sc XIII}). This may again be reflective of different SFHs, as galaxies with predominantly red HBs are expected to be deficient in short-period, high-amplitude pulsators and could manifest a narrower RRab period distribution than galaxies with more extended HBs. Peculiarities in the chemical enrichment history can also play a role. RRab stars of different metallicity are known to follow different period-amplitude loci, with metal-poor RR Lyrae being shifted to longer periods at a given amplitude\citep[e.g.,][]{Clementini22}.
Therefore, potential correlations between the metal abundance and pulsation amplitude, produced by a peculiar star formation and chemical enrichment history, could naturally alter the morphology of the period-amplitude sequence. 

Of course the populations shown in Fig.~\ref{fig:Bailey} are also affected by observational effects. In the case of the less populated Bailey diagrams, the inevitable stochasticity in the RR Lyrae population may introduce apparent differences, as the position of only a handful of stars can easily affect the Bailey diagram morphology. Another important effect is the limited coverage of the ACS field of view, that reduces the number of detected RR Lyrae. This is particularly relevant in galaxies with large apparent sizes. The prime example of this effect (besides the obvious cases of M31 and M33) is the strongly elongated And~{\sc XIX} for which, due to the half-light radius of 14.2\arcmin\ \citep{Martin16}, we detected only four RR Lyrae in spite of the relatively bright luminosity of this galaxy ($M_V=-10.1\pm0.3$). The effect of a limited field of view, in the presence of spatial gradients in the stellar population properties, can also alter the measured RRab/RRc ratio. Such field-of-view effects, however, become less pronounced for fainter galaxies. For the smaller dwarfs, our ACS field generally covers 2-3 half-light radii and differences in the RR Lyrae demographic of galaxies with comparable luminosity (such as And~{\sc XVI} versus And~{\sc X}, or And~{\sc XII} versus And~{\sc XXII}) become representative of intrinsic differences in their stellar populations.

Because of the interplay among these numerous astrophysical and observational effects, the detailed interpretation of our Bailey diagrams requires a quantitative modeling framework that we will explore once we have measured star formation histories for our full sample of galaxies (Savino et al.\ in prep).

\subsection{Quantifying the Incidence of Period Aliases}
The analysis procedure described in \S~\ref{sec:RRmod} was designed to maximize fit robustness on sparsely-sampled light curves. Even so, retrieving accurate pulsation properties of variable stars with few photometric observations remains a challenging task. The most common problem that can arise in a low light-curve sampling regime is the contamination from period aliases. These are spurious variability signals that arise from the discrete nature of the photometric observations and, if strong enough, can lead to an incorrect period being assigned to the star. A high enough incidence of period aliases, would therefore be detrimental for our distance determination accuracy.

Quantification of period alias incidence is usually done by applying the variable star analysis to light curves of known period. In our case, we leveraged the large range of observation cadence in our sample galaxies and used stars with highly-sampled light curves as templates, degrading them to simulate the range of observing conditions in our most sparsely sampled galaxies. Specifically, we selected three variable stars belonging to M31, chosen to represent different light curve properties: a high-amplitude RRab star (V005030, $P=0.559~d$, $A_{F606W}=1.008$), a low-amplitude RRab star (V005674, $P=0.734~d$, $A_{F606W}=0.436$), and an RRc star (V005402, $P=0.283~d$, $A_{F606W}=0.499$). We took the well-measured light curves of these stars (58 epochs in F606W and 60 epochs in F814W) and selected random subsets of data with variable length, ranging from 9 to 20 photometric epochs per filter. The photometric time-series were selected to span a small range in observation epoch and without any significant time gap, to simulate typical \hst\ data. For each number of photometric epochs, we generated 200 random light-curve realizations.

We then analyzed the simulated light curves as we did with our observed sample, following the procedure of \S~\ref{sec:RRmod}, and quantified the amount of solutions affected by a period alias. Within this experiment, we defined a solution to be affected by an alias if the recovered logarithm of the period differs from the real one (i) in excess of measurement  uncertainties and (ii) by more than 0.05 dex. We chose this value as it represents a period systematic that is comparable with the other sources of uncertainties in our distance determination. Aliases at smaller period separations may still be possible, however they would not have any significant impact on the results of this paper.

\begin{figure}

\plotone{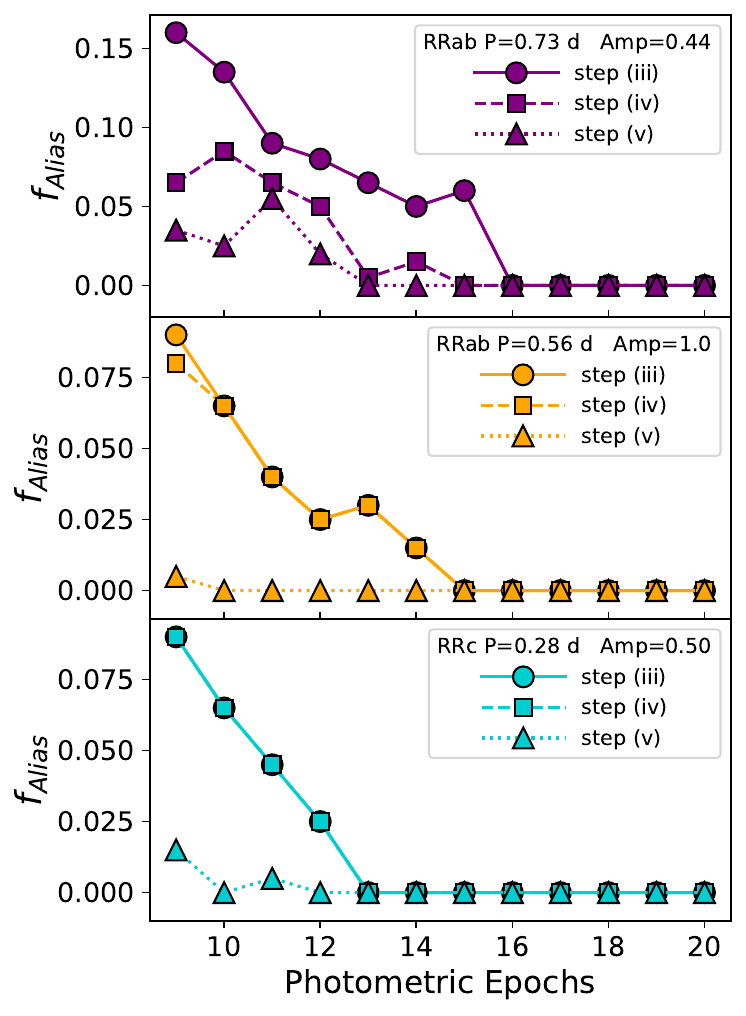}
\caption{Fraction of solutions affected by period aliases, as a function of the number of time-series photometric epochs per filter, at the end of step (iii, circles), step (iv, squares) and step (v, triangles) of our pipeline. Results are showed for the simulated light curves of a long-period RRab star (top panel), a short-period high-amplitude RRab star (middle panel) and an RRc star (bottom panel).}
\label{fig:alias}
\end{figure}

Figure~\ref{fig:alias} shows the fraction of simulated light curves whose solution is affected by a period alias, as a function of the number of photometric epochs per filter, at different stages of our analysis pipeline. For light-curve samplings greater than approximately 15 epochs per filter, we efficiently recover the correct pulsation period, already during our initial guess (step iii, circles in Fig.~\ref{fig:alias}). This highlights the robustness of the \citet{Saha17} methodology, which was specifically developed to minimize period ambiguity in sparsely populated light curves. At lower sampling values, aliases in the first-guess period start increasing in frequency, reaching an incidence between 5\% and 15\% in the sparsest sampling regime of our data (9 to 11 epochs per filter). This fraction is only marginally reduced by the template fitting analysis (step iv, squares), as the algorithm preferentially samples the parameter space in the vicinity of the initial period. Many of the spurious solutions however, result in period-amplitude combinations that are in contrast to what expected for RR Lyrae and they are therefore flagged as anomalous (step v, triangles). This post-processing step drastically reduces the incidence of period aliases to values below 5\%, at all samplings. The efficiency of step (v) in identifying aliases is in agreement with results from our real sample. In fact, out of the $\sim 5000$ RR Lyrae candidates processed in \S~\ref{sec:RRmod}, roughly 10\% were flagged in step (v). Of these, approximately 75\% were flagged because of an evident period alias, in line with the decrease observed, in Fig.~\ref{fig:alias}, between step (iv) and step (v).

The prevalence of aliases seems to be more important for long-period, low-amplitude RRab stars, for which it occurs at roughly double the rate of the other variable templates. This is reasonable as, at fixed light-curve sampling, observations will cover fewer pulsation cycles (or even just a part of the pulsation cycle) in long-period variables, therefore increasing the chance of misclassification. Furthermore, high-amplitude pulsators are known to exist predominantly as short-period RRab stars, and will manifest clearly unphysical aliased solutions (either high-amplitude RRc solutions or long-period high-amplitude RRab). In constrast, spurious solutions in long-period RRab stars could potentially manifest as low-amplitude RRc light-curves and therefore the alias would be less easily identified.

From the results of Fig.~\ref{fig:alias}, we estimated that the fraction of unidentified period aliases in our final sample is likely no larger than a few percent and mostly prevalent among long-period RRab stars. This low fraction of inaccurate solutions effectively act as a contaminating population in our distance fit and, given the low prevalence, is efficiently taken into account by the GMM formalism described in \S~\ref{sec:Model}.

\section{Distance determination} \label{sec:distances}
\subsection{Defining the Period-Wesenheit Magnitude - Metallicity Relationship}
\label{sec:PWZ}

To derive robust distances to our sample, we made use of our multi-band photometry to construct Wesenheit magnitudes \citep{Madore82}, defined as:
\begin{subequations}
    \begin{align}
        &W(X,X-Y) = X-R(X-Y)\\
        &R=A_X/E(X-Y),
    \end{align}
    \label{eq:W}
    \end{subequations}
where $R$ is the total-to-selective dust absorption ratio for the $X,Y$ filter pair. Wesenheit magnitudes, by construction, have the advantage to be reddening-free and are widely used as distance indicators, with both RR Lyrae and Cepheids \citep[e.g.,][]{Neeley19,Riess21}. They are known to be tightly correlated with period and metallicity of the variable star, according to the following functional form:
\begin{equation}
    W=\mu+a+b\, \log{P} +c\, {\rm[Fe/H]},
    \label{eq:PWZ}
\end{equation}
where $\mu$ is the distance modulus to the star.


\begin{table*}
    
    \caption{Coefficient of the ACS/WFC PWZ relations used in this paper (see Eq.~\ref{eq:PWZ}), obtained from the models of \citet{Marconi15}.}
    \centering
    \begin{tabular}{llrrrr}

    \toprule
    Pulsation Mode&Filters&$R$&$a$&$b$&$c$\\
    \toprule
    RRab&W(F814W,F475W-F814W)&0.960&-0.990$\pm$0.007&-2.394$\pm$0.025&0.129$\pm$0.004\\
    RRc&W(F814W,F475W-F814W)&0.960&-1.390$\pm$0.012&-2.490$\pm$0.026&0.119$\pm$0.004\\
    RRab&W(F814W,F606W-F814W)&1.785&-0.966$\pm$0.006&-2.374$\pm$0.022&0.157$\pm$0.004\\
    RRc&W(F814W,F606W-F814W)&1.785&-1.381$\pm$0.010&-2.518$\pm$0.023&0.140$\pm$0.004\\
    \toprule

    \end{tabular}
    \label{tab:PWZ}
    
\end{table*}

With the general form of the RR Lyrae period-Wesenheit-metallicity (hereafter PWZ) in hand, we must decide on what distance anchor to use: empirical, theoretical, or a mixture of the two \citep[e.g.,][]{Kovacs97,Braga15,Marconi15,Neeley19,Marconi21}.  Empirical calibrations have the advantage that the RR Lyrae absolute W magnitude can be tied to independent distances (e.g., via geometric parallax, through cluster CMD fitting). A particularly relevant example is that of \citet[hereafter N21]{Nagarajan22}, which used a sample of 36 local RR Lyrae stars with excellent ground-based light curves and metallicities to empirically calibrate a PWZ relation anchored to \textit{Gaia}\ eDR3 parallaxes.  This is among the first to tie a RR Lyrae-based population II distance indicator to the \textit{Gaia}\ eDR3 reference frame.

However, empirical anchors also have limitations, particularly for the work at hand.  To date, PWZ optical calibrations essentially only exist in ground-based filters (e.g., Johnson B, V, I) or for \textit{Gaia}\ magnitudes, whereas our data is taken in the native \hst\ photometric system. Transforming magnitudes between different photometric systems is notoriously difficult, particularly when the two systems are sufficiently different from each other. Though some \hst\ and Johnson filters are considered similar, a close inspection of available \hst-Johnson filter transformations present in literature shows that systematic differences of up to 0.1 mag can arise between different empirical and/or synthetic transformations \citep{Sirianni05,Saha11,Bernard09,McQuinn15}. Such differences are amplified by the use of W magnitudes, which are defined as a linear combination of different filters. Adopting $R=0.960$ (F475W/F814W), a systematic of 0.1 mag in one of the two filters could translate to a difference in W of up to $\sim$ 0.2 mag, while a value of $R=1.785$ (F606W/F814W), could result in systematics up to the $\sim$ 0.3 mag level. As a reference, our median uncertainty on W is 0.11 mag. As \citet{Riess21} note, ground-to-\hst\ filter transformations are one of the dominant sources of uncertainty on galaxy distances in the limit of excellent data, such as ours.

In comparison, theoretical PWZ calibrations provide a way to mitigate this problem. Bolometric luminosities of model RR Lyrae can be readily translated into any photometric system of choice, eliminating the systematics inherent to empirical transformations. One of the most widely used theoretical calibrations is provided by \citet[hereafter M15]{Marconi15}, who use a grid of non-linear pulsation models to explore the dependency of W magnitudes on a large range of stellar mass, period, and metallicity. Such calibrations, opportunely mapped onto the ACS/WFC photometric system (M. Marconi, private communication), constitute the basis of our distance determinations. The PWZ coefficients we used in this work are listed in Tab.~\ref{tab:PWZ}. Our distances were determined using only data and PWZ coefficients for the RRab stars. In \S~\ref{sec:RRc} we will examine the effect of this choice and discuss modeling of RRc stars.

A drawback of theoretical calibrations is that they are less easily informed by independent distance anchors, resulting in potentially substantial differences from what is measured through, e.g., parallaxes. The work from \hyperlink{cite.Nagarajan22}{N21}, for instance, reports that the \hyperlink{cite.Marconi15}{M15} PWZ systematically overpredicts distances compared to \textit{Gaia}\ eDR3. This is easily appreciated by comparing the PWZ coefficients that both authors provide for the W(I,V-I)-based PWZ. As the transformations to convert bolometric luminosities into Johnson magnitudes are among the most well-established, it is reasonable that the dominant source in the predicted W(I,V-I) difference, for a given period and metallicity, lies in the absolute luminosity of the theoretical pulsation models. 

Under this assumption, we derived an empirical correction term, defined as the difference between the W(I,V-I)-based PWZ of \hyperlink{cite.Marconi15}{M15} and that of \hyperlink{cite.Nagarajan22}{N21}, and used it to recalibrate the \hst\ PWZs to be consistent with the \textit{Gaia}\ eDR3 distance scale. The functional form for this correction term is:

\begin{equation}
\begin{split}
    \Delta\mu=\,&0.23 \log{P}-0.02 [Fe/H]\\
    &-0.087 (V-I).
\end{split}
\label{eq:Gaia}
\end{equation}

\begin{figure}

\plotone{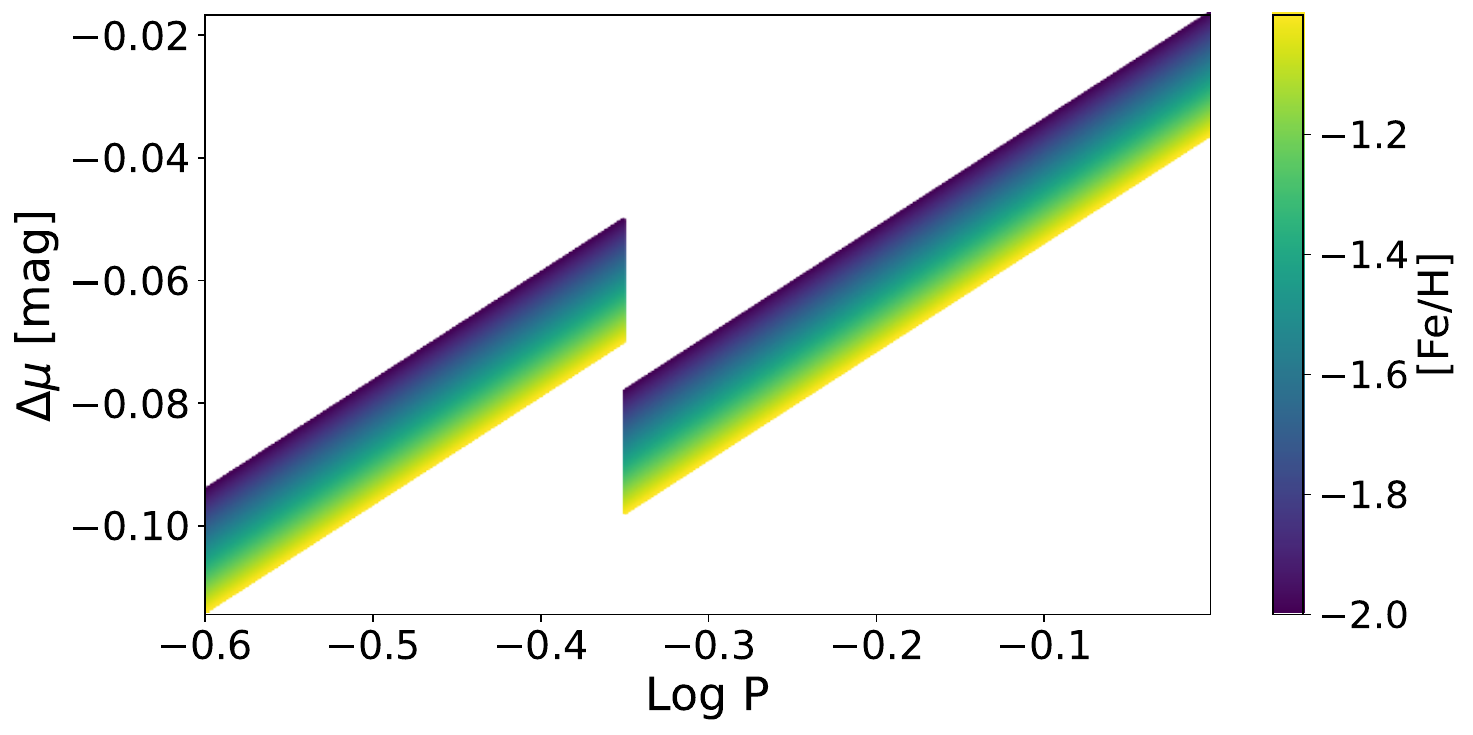}
\caption{Empirical correction (from Eq.~\ref{eq:Gaia}) to the distance obtained with the \citet{Marconi15} PWZ, shown as function of period and metallicity, assuming a linear correlation between color and period. For RRab stars ($\rm \log{P} > -0.35$), distances anchored to \textit{Gaia}\ eDR3 parallaxes are 0.02-0.1 mag closer than those on the original \citet{Marconi15} distance scale.}
\label{fig:Correction}
\end{figure}

Figure~\ref{fig:Correction} illustrates the overall size of the empirical correction term as a function of period and [Fe/H]. For periods typical of RRab pulsators ($\rm -0.35<\log{P}<0$), $\mu$ decreases by 0.02 to 0.1~mag, in line with the findings of \hyperlink{cite.Nagarajan22}{N21}. For the illustrative purposes of Fig.~\ref{fig:Correction}, we have assumed a linear correlation between the period and the ($V-I$) color, so that a period range of $\rm -0.6<\log{P}<0$ would map onto the color range $0.3<(V-I)<0.6$. The color term in Eq.~\ref{eq:Gaia} arises from the fact that \hyperlink{cite.Marconi15}{M15} and \hyperlink{cite.Nagarajan22}{N21} use different values of $R$ in their definition of W($I,V-I$). To calculate this color term in Eq.~\ref{eq:Gaia}, we transformed the intensity-averaged F606W/F814W magnitudes to V/I using the prescription of \citet{Saha11} and the F475W/F814W magnitudes to V/I using that of \citet{McQuinn15}. The presence of this color term implies that a transformation from \hst\ magnitudes to Johnson was still involved in our procedure, which was otherwise performed entirely in the native HST photometric system. However, the small coefficient of this color term means that any systematic arising from the filter transformations affects the final distance only by a factor of order 0.01~mag. Similarly, the quoted random uncertainties in the filter transformations only contribute a term of order 0.003~mag to the final distance error budget.

\begin{table}
\centering
\caption{Parameters and priors used in our PWZ modeling.}
\begin{tabular}{lll}


\toprule
Parameter&Prior&Description\\   
\toprule
$\mu$&$\mathcal{U}(23,26)$&Distance modulus of the galaxy\\
$[Fe/H]$&$\mathcal{U}(-2,-1)$&Metallicity of the RR Lyrae population\\
$\mu_{false}$&$\mathcal{U}(20,30)$&Mean distance modulus of the contaminants\\
$\sigma_{false}$&$\mathcal{U}(0,10)$&Scatter of the contaminant population\\
$s$&$\mathcal{U}(1,10)$&Steepness of the $Q$ sigmoid\\

\toprule

\end{tabular}

\label{tab:PWZprior}
\end{table}

\begin{figure}

     \subfloat
	{\includegraphics[width=0.45\textwidth]{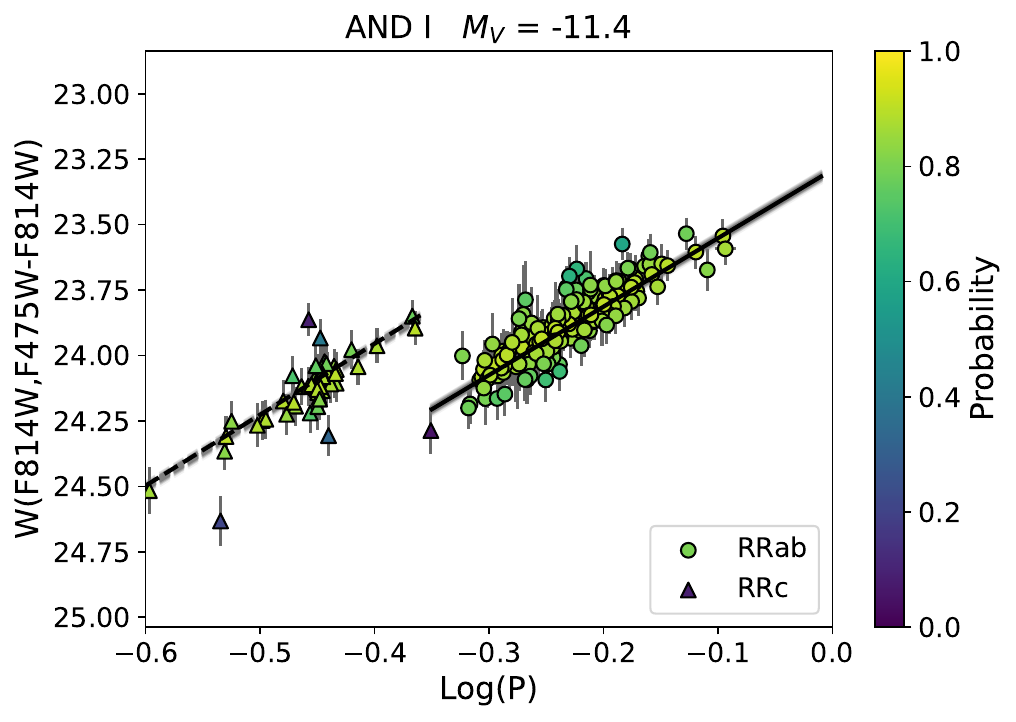} \label{fig:PWZA1}} \quad
	 \subfloat
	{\includegraphics[width=0.45\textwidth]{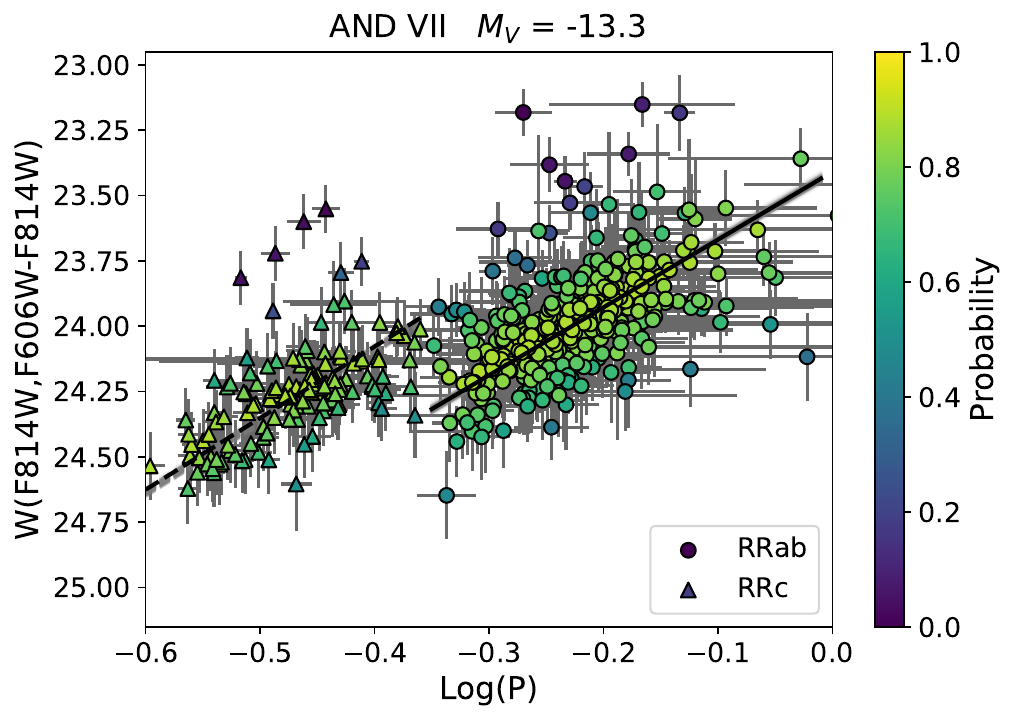} \label{fig:PWZA7}} \quad
	\subfloat
	{\includegraphics[width=0.45\textwidth]{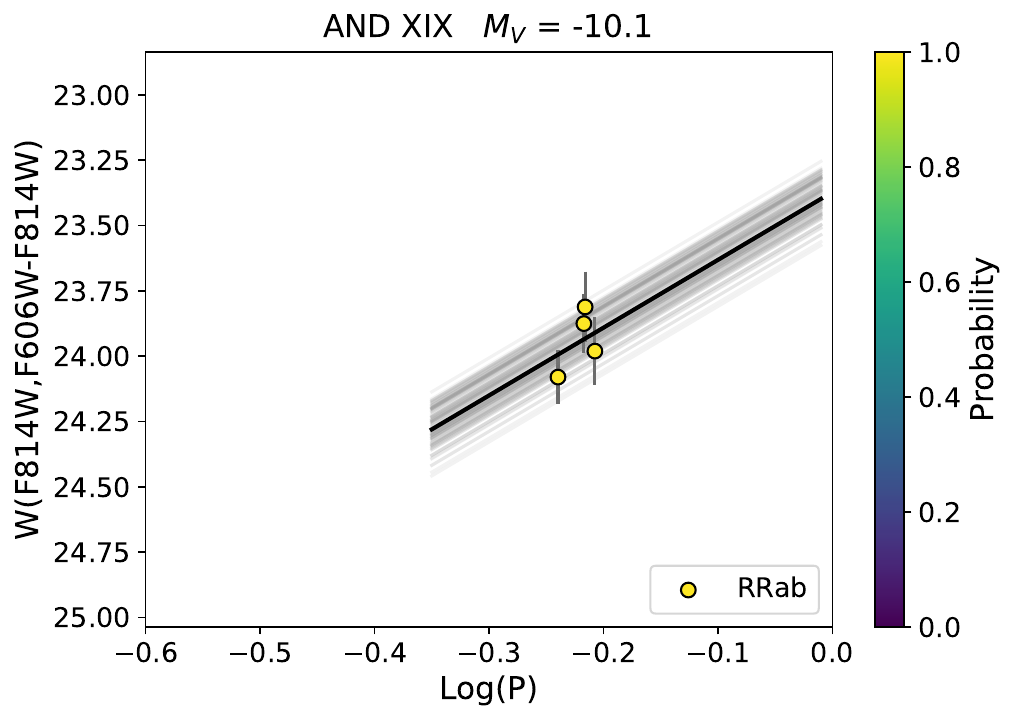} \label{fig:PWZA19}} \quad
\caption{Representative observed PWZ for the galaxies in our sample, showing different data quality regimes: a populous, well-characterized RR Lyrae sample (And~{\sc I}, top panel); a populous, noisy sample (And~{\sc VII}, center); and a sparsely populated sample (And~{\sc XIX}, bottom). Circles and triangles represent RRab and RRc, respectively, and are color coded by the probability of being bona-fide measurements, ($1-Q_k$). The solid and dashed lines show our best-fit model for RRab and RRc respectively. Grey lines show 100 random PWZ realizations from the MCMC chains.}
\label{fig:PWZ}
\end{figure}

\subsection{Measuring Galaxy Distances}
\label{sec:Model}

We measured the distance to each galaxy using the periods and mean magnitudes of individual RR Lyrae as determined in \S \ref{sec:variables}, the PWZ from \S \ref{sec:PWZ}, and a GMM formalism, which provides a way to account for the effect of outlier RR Lyrae stars on the galaxy distance without resorting to hard cuts (e.g., sigma-clipping). We followed the formalism for a GMM as laid out in \citet{Hogg10} and \citet{Foreman-Mackey14}.


For each RR Lyrae in a given galaxy, we used the measured mean magnitudes and period to calculate a distance modulus $\mu_k$ by applying Eq.~\ref{eq:W},~\ref{eq:PWZ}, and \ref{eq:Gaia}. We assumed the measured $\mu_k$ to be sampled from a normal distribution $\mathcal{N}(\mu,\sigma_k^{\mu})$, where $\mu$ is the true distance modulus of our galaxy and $\sigma_k^{\mu}$ is determined, for each star, through propagation of the measurement uncertainties (P and W) and of the uncertainties in the PWZ coefficients. The metallicity of each RR Lyrae, needed to use Eq.~\ref{eq:PWZ} and \ref{eq:Gaia}, is not known (i.e., they are too faint for metallicity determinations), so we left it as a free parameter, with the assumption that the value is the same for all RR Lyrae in the galaxy. 

Despite our extensive efforts to remove contaminants from our RR Lyrae sample, our sample may not be 100\% pure.  For example, stars with few epochs could have incorrect periods due to aliasing or could be non-RR Lyrae that are misidentified, both of which would introduce contamination into the sample, possibly affecting the distance determinations.


To account for contamination, we adopted a GMM to model a second population which is drawn from $\mathcal{N}(\mu_{false},\sigma_{false})$, where $\mu_{false}$ and $\sigma_{false}$ are both free nuisance parameters of our model. We did not enforce a binary decision on whether a given measurement belonged to either population, but rather assigned to each star a continuous probability, $ Q_k$, of being a contaminant. We modeled the probability function as a sigmoid-like function:
\begin{equation}
Q_k = \frac{1}{{1+exp({-s(R_k-2)}})},
\end{equation}
where $R_k = |\mu_k - \overline{\mu}|/\sigma_k^{\mu}$ is the absolute deviation, expressed in units of $\sigma_k^{\mu}$, from the weighted average of the $\mu_k$ distribution. We chose a sigmoid so that measurements in the vicinity of the observed Period-Wesenheit sequence contribute fully to the fit, while scattered outliers have significantly reduced constraining power.  The probability $Q_k$ is defined to be equal to 0.5 at $R_k = 2$ and to approach 1 as $R_k$ tends to infinity. The parameter $s$ sets the slope of the sigmoid cut-off, and it was left free to accommodate galaxy-by-galaxy differences in data quality and contamination level. Within this framework, the likelihood of a given parameter combination can be written as:

\begin{equation}
    \begin{split}
    p(\mu_k | \mu, [Fe/H],  \mu_{false}, \sigma_{false}, s)= \\
    \\
    (1-Q_k)\, exp(-(\mu_k-\mu)^2/2(\sigma_k^{\mu})^2) \\
    \\
    + Q_k\,  exp(-(\mu_k-\mu_{false})^2/2(\sigma_{false})^2).
\end{split}
\label{eq:likelihood}
\end{equation}

We explored this parameter space with \texttt{emcee}, using the same convergence criterion defined in \S~\ref{sec:RRmod}, the sum of the log-likelihoods of all RRab stars as our merit function, and flat, uninformative priors on our parameters. Specifically, we defined the priors as a top-hat function, whose limits are listed in Tab.~\ref{tab:PWZprior}. For $\mu, \mu_{false}, \sigma_{false}$ and $s$, we chose the width of the top-hat to be large enough to accommodate extensive exploration of our MCMC walkers. The prior on [Fe/H] is flat for $\rm -2<[Fe/H]<-1$ and zero outside of this interval. The reason for this choice is that our data have little to no informative power on [Fe/H], meaning that, in virtually every fit, the recovered PPD on [Fe/H] closely tracks the prior. Choosing a top-hat prior means that we effectively marginalized over a flat metallicity posterior, while ensuring that [Fe/H] remains within a physically motivated interval. This essentially captures the effect of our ignorance on the metallicity in the final distance uncertainties. We chose the limits of [Fe/H] to be representative of the expected metallicity range predicted by stellar evolution theory \citep[e.g.,][]{Savino20} and observed in local dwarf galaxies \citep[e.g.,][]{Clementini05,Bernard08}.

\begin{table*}
    
    \caption{Our measured distance moduli and the resulting distances relative to both the Sun and M31.  The values reported are the best fit to the RRab variables.  We list the number of RRab and RRc we identify in each system. For the satellite galaxies, we indicate the probability $Q_{Plane}$ of belonging to the planar structure identified in \S~\ref{sec:plane}. We also report updated absolute luminosities and physical half-light radii for the dwarf galaxies, using the new distances. The references for the apparent V magnitudes and apparent sizes are also listed. For M32, NGC~147, NGC~185, and NGC~205, the reported radius is not the half-light radius, but rather the effective radius of a S\'ersic profile with index: 4 (M32 and NGC 205), 1.69 (NGC 147) and 1.78 (NGC 185).}
    \centering
    \begin{tabular}{lllcrrrrrr}
    \toprule
         ID & $N_{ab}$&$N_c$&$\mu$ &$D_{\odot}$& $D_{M31}$ & $Q_{Plane}$&$M_V$&$r_h$ & Ref. \\
            &    &&       & kpc& kpc & &&pc& \\
         \toprule
         M31&28&21&$24.45\pm 0.06$&776.2$_{-21}^{+22}$&$0$&...&...&...&...\\
         GSS&16&5&$24.58\pm 0.07$ & $824.1_{-26}^{+27}$&$53.4^{+30}_{-26}$&...&...&...&...\\
         M32&145&49&$24.44\pm 0.06$&$772.7_{-21}^{+22}$&$20.6^{+21}_{-13}$&0.953&$-16.8\pm0.1$&$106\pm12$&(1)\\
         M33&47&10&$24.67\pm 0.06$&$859.0_{-23}^{+24}$&$226.7^{+15}_{-11}$&$\sim 0$ &...&...&...\\
         NGC147&90&34&$24.33\pm 0.06$&$734.5_{-20}^{+21}$&$107.0^{+15}_{-8.0}$&0.856&$-16.6\pm0.07$&$1431_{-43}^{+44}$&(2)\\
         NGC185&387&192&$24.06\pm 0.06$&$648.6\pm18$&$154.1^{+23}_{-21}$&0.931&$-15.6\pm0.07$&$555\pm17$&(2)\\
         NGC205&261&145&$24.61\pm 0.06$&$835.6\pm23$&$58.0^{+30}_{-29}$&0.919&$-16.7\pm0.1$&$598_{-29}^{+30}$&(1)\\
         And~{\sc I}&181&53&$24.45\pm 0.05$&$776.2\pm18$&$48.0^{+10}_{-3.2}$&0.980&$-11.4\pm0.2$&$880_{-30}^{+31}$&(3)\\
         And~{\sc II}&152&51&$24.12\pm 0.05$&$666.8_{-15}^{+16}$&$168.9^{+19}_{-16}$&0.017&$-11.7\pm0.2$&$1028_{-30}^{+31}$&(3)\\
         And~{\sc III}&82&23&$24.29\pm 0.05$&$721.1_{-16}^{+17}$&$84.9^{+19}_{-14}$&0.188&$-9.5\pm0.2$&$420\pm43$&(3)\\
         And~{\sc V}&122&39&$24.40\pm 0.06$&$758.6\pm21$&$110.5^{+7.0}_{-3.5}$&$0.002$&$-9.3\pm0.2$&$353_{-24}^{+35}$&(3)\\
         And~{\sc VI}&203&55&$24.60\pm 0.06$&$831.8\pm23$&$281.6^{+8.6}_{-7.1}$&$\sim 0$&$-11.6\pm0.2$&$489\pm22$&(4)\\
         And~{\sc VII}&418&145&$24.58\pm 0.06$&$824.1\pm23$&$230.8^{+8.4}_{-6.5}$&$\sim 0$&$-13.3\pm0.3$&$815\pm28$&(4)\\
         And~{\sc IX}&41&11&$24.23\pm 0.06$&$701.5_{-19}^{+20}$&$82.0^{+26}_{-24}$&0.339&$-8.6\pm0.3$&$408_{-42}^{+62}$&(3)\\
         And~{\sc X}&48&16&$24.00\pm 0.06$&$631.0_{-17}^{+18}$&$162.2^{+25}_{-24}$&0.057&$-7.3\pm0.3$&$202_{-37}^{+74}$&(3)\\
         And~{\sc XI}&15&9&$24.38\pm 0.07$&$751.6_{-24}^{+25}$&$104.2^{+11}_{-4.2}$&0.669&$-6.4\pm0.4$&$131\pm44$&(3)\\
         And~{\sc XII}&7&5&$24.28^{+0.08}_{-0.07}$&$717.8_{-23}^{+27}$&$107.7^{+20}_{-13}$&0.762&$-6.6\pm0.5$&$376_{-147}^{+251}$&(3)\\
         And~{\sc XIII}&11&3&$24.57\pm 0.07$&$820.4_{-26}^{+27}$&$126.4^{+16}_{-8.0}$&0.803&$-6.8\pm0.4$&$191_{-72}^{+96}$&(3)\\
         And~{\sc XIV}&90&22&$24.44\pm 0.06$&$772.7_{-21}^{+22}$&$160.8^{+3.8}_{-4.2}$&0.835&$-8.6\pm0.3$&$337\pm46$&(3)\\
         And~{\sc XV}&75&34&$24.37\pm 0.05$&$748.2\pm17$&$95.8^{+12}_{-4.8}$&0.004&$-8.4\pm0.3$&$283\pm23$&(3)\\
         And~{\sc XVI}&3&4&$23.57\pm 0.08$&$517.6\pm19$&$280.0^{+26}_{-27}$&0.874&$-7.5\pm0.3$&$239\pm25$&(3)\\
         And~{\sc XVII}&27&20&$24.40\pm 0.07$&$758.6_{-24}^{+25}$&$49.9^{+17}_{-5.8}$&0.798&$-7.8\pm0.3$&$315\pm68$&(3)\\
         And~{\sc XIX}&4&0&$24.55^{+0.09}_{-0.08}$&$812.8_{-29}^{+34}$&$113.3^{+18}_{-6.9}$&0.001&$-10.1\pm0.3$&$3357_{-465}^{+816}$&(3)\\
         And~{\sc XX}&12&3&$24.35\pm 0.08$&$741.3_{-27}^{+28}$&$128.4^{+12}_{-5.5}$&$\sim 0$&$-6.4\pm0.4$&$86_{-22}^{+43}$&(3)\\
         And~{\sc XXI}&21&2&$24.44^{+0.06}_{-0.07}$&$772.7_{-25}^{+22}$&$124.4^{+5.1}_{-3.8}$&$\sim 0$&$-8.9\pm0.3$&$922_{-95}^{+182}$&(3)\\
         And~{\sc XXII}&16&4&$24.39\pm 0.07$&$755.1_{-24}^{+25}$&$216.8^{+5.7}_{-5.6}$&$\sim 0$&$-6.4\pm0.4$&$198_{-44}^{+66}$&(3)\\
         And~{\sc XXIII}&15&3&$24.36\pm 0.07$&$744.7\pm24$&$128.1^{+10}_{-4.9}$&$\sim 0$ & $-9.8\pm0.2$ & $1170_{-94}^{+95}$ &(3)\\
         And~{\sc XXIV}&16&5&$23.92\pm 0.07$&$608.1_{-19}^{+20}$&$194.5^{+25}_{-24}$&0.006&$-7.6\pm0.3$&$460_{-90}^{+178}$&(3)\\
         And~{\sc XXV}&18&9&$24.38^{+0.07}_{-0.06}$&$751.6_{-21}^{+25}$&$85.2^{+12}_{-4.4}$&0.553&$-9.1_{-0.2}^{+0.3}$&$590_{-47}^{+90}$&(3)\\
         And~{\sc XXVI}&21&9&$24.48^{+0.06}_{-0.07}$&$787.0_{-25}^{+22}$&$104.6^{+6.8}_{-3.5}$&0.312&$-6.0_{-0.5}^{+0.7}$&$229_{-115}^{+138}$&(3)\\
         And~{\sc XXVIII}&36&43&$24.36\pm 0.05$&$744.7\pm17$&$368.8^{+7.8}_{-7.3}$&$\sim 0$&$-8.8_{-1.0}^{+0.4}$&$240\pm46$&(5)\\
         And~{\sc XXIX}&45&10&$24.26\pm 0.06$&$711.2_{-19}^{+20}$&$189.1^{+12}_{-8.8}$&$\sim 0$&$-8.2\pm0.4$&$352_{-42}^{+43}$&(6)\\
         And~{\sc XXX}&37&15&$23.74\pm 0.06$&$559.8_{-15}^{+16}$&$238.6^{+24}_{-24}$&0.855&$-7.7_{-0.2}^{+0.3}$&$245\pm33$&(3)\\
         And~{\sc XXXI}&42&15&$24.36\pm 0.05$&$744.7\pm17$&$261.4^{+6.9}_{-5.9}$&$\sim 0$&$-11.6\pm0.7$&$910_{-110}^{+89}$&(7)\\
         And~{\sc }XXXII&71&14&$24.52\pm 0.06$&$801.7_{-22}^{+23}$&$146.8^{+7.8}_{-4.2}$&0.582&$-12.3\pm0.7$&$1516_{-237}^{+283}$&(7)\\
         And~{\sc XXXIII}&35&6&$24.24\pm 0.06$&$704.7_{-19}^{+20}$&$340.3^{+10}_{-8.7}$&$\sim 0$&$-10.1\pm0.7$&$348\pm83$&(8)\\
         Pisces &52&8&$23.91\pm 0.05$&$605.3\pm14$&$292.1^{+17}_{-16}$&0.900&$-9.8\pm0.1$&$370\pm36$&(9)\\
         Peg DIG&530&182&$24.74\pm 0.05$&$887.2_{-20}^{+21}$&$458.2^{+11}_{-9.4}$&...&$-12.3\pm0.2$&...&(10)\\
         IC 1613&58&23&$24.32\pm 0.05$&$731.1\pm17$&$511.1^{+10}_{-9.8}$&...&$-15.2\pm0.1$&...&(10)\\
         \toprule
    \end{tabular}
    \begin{tablenotes}
    \item \textbf{References}: (1) \citet{Choi02}; (2) \citet{Crnojevic14}; (3) \citet{Martin16}; (4) \citet{McConnachie06b}; (5) \citet{Slater11}; (6) \citet{Bell11}; (7) \citet{Martin13a}; (8) \citet{Martin13b}; (9) \citet{Lee95}; (10) \citet{McConnachie12}.
    \end{tablenotes}
    \label{tab:DM}
\end{table*}

Figure~\ref{fig:PWZ} shows examples from our PWZ models, illustrating different regimes of data quality within our sample: a clean, populated sample (Fig.~\ref{fig:PWZA1}, And~{\sc I}); a populated, noisy sample (Fig.~\ref{fig:PWZA7}, And~{\sc VII}); and a sparsely populated sample (Fig.~\ref{fig:PWZA19}, And~{\sc XIX}). However, the large numbers of RR Lyrae mean that we are able to achieve robust PWZ fits even in the presence of a modest population of contaminants or uncertain pulsation properties on individual stars, consistent with expectations for simple model constraints from noisy data \citep[e.g.,][]{Hogg10}. The distance moduli measured through our PWZ models (50th percentile of the PPD), and the respective uncertainties (15.9th and 84.1th percentiles of the PPD), are summarized in Tab.~\ref{tab:DM}. 

For most of our targets, the dominant source of random uncertainties is the unknown metallicity of the RR Lyrae, which limits precision to roughly 0.05 magnitudes. Uncertainties in the variability parameters and small number of RR Lyrae stars are only a significant contribution to the distance uncertainties of the least populated ($N_{ab}\lesssim 15$) galaxies, such as And~{\sc XIX}. This is due to our choice of incorporating a broad [Fe/H] prior in our modeling, so that the random uncertainties resulting from our fits take into account the precision of our pulsation parameters, the PWZ uncertainties reported in Tab.~\ref{tab:PWZ}, as well as the poorly-constrained metallicity of the RR Lyrae stars.

\subsection{The Distance to M31}
\label{sec:M31}

At the heart of determining the 3D structure of the M31 satellite system is a self-consistent distance to M31 itself. From our re-reduction of archival M31 \hst\ imaging, we apply our PWZ and distance modeling to 28 RRab variables to find $\mu=24.45\pm0.06$, which corresponds to a physical distance of $776.2^{+22}_{-21}$~kpc. Previous analysis of the RR Lyrae stars in the same field resulted in $\mu=24.48\pm0.15$  \citep{Jeffery11}.  Though this field is located in the inner stellar halo at a projected distance of 11~kpc from the photometric center of M31, under the approximation of reasonable spherical symmetry for metal-poor halo RR Lyrae, our distance value is virtually identical to the distance to M31's center.



Recent literature values of the M31 distance obtained from Cepheids are $\mu=24.41\pm0.03$ \citep{Li21} and $\mu=24.46\pm0.20$ \citep{Bhardwaj16}, while eclipsing binary studies yield $\mu=24.44\pm0.12$ \citep{Ribas05} and $\mu=24.36\pm0.08$ \citep{Vilardell10}. From a meta-analysis of 34 literature measurements, \citet{deGrijs14} derive a M31 distance modulus of $24.46\pm0.1$. Our measured value is in good agreement with these results and has been derived consistently with the distance to the other target in our sample. Therefore we adopted the distance of $\mu=24.45\pm0.06$ as our anchor point for the M31 system and use it to derive relative distances to M31 (listed in Tab.~\ref{tab:DM}) and the 3D structure of our sample. We detail the procedure for deriving 3D physical distances in \S~\ref{sec:Cartesian}.

\subsection{Revised Luminosities and Physical Sizes}
The structure and morphology of the M31 satellites have been studied in great detail over past decades \citep[e.g.,][]{Choi02,McConnachie06, McConnachie12}. As more satellites have been discovered, a large effort to uniformly derive structural parameters was undertaken by the PAndAS team \citep[e.g.,][]{Crnojevic14,Martin16}. At the level of the full M31 satellite system, however, absolute luminosities and physical sizes existing in the literature are still somewhat heterogeneous, with some of the pre-PAndAS determinations being based on a range of distances and extinctions. This makes population-wide comparisons challenging. 

As one step toward homogenizing the physical properties of M31 satellites, we use our new distance catalog to update some of the structural parameters (i.e., sizes and luminosities) for our galaxy sample. Though in general, distances to most galaxies do not change dramatically, there are a handful of cases in which we do see large changes compared to values used for computing sizes and luminosities (e.g., And~{\sc IX}, And~{\sc XV}, and And~{\sc XXIV}). 

To update sizes and luminosities, we took apparent magnitudes and angular sizes from a set of literature studies chosen to be as homogeneous as possible (references given in Tab.~\ref{tab:DM}) and converted them into absolute V luminosities and physical half-light radii using our RR-Lyrae distances and extinction values from \citet{Schlafly11}. We report the updated luminosities and sizes, the first homogeneously derived for virtually the full M31 system, in Tab.~\ref{tab:DM}. For most galaxies, the luminosities and sizes change by a modest amount, comparable to or smaller than the measurement errors, with respect to literature values. The only galaxy for which we find substantial differences is And~{\sc XXIV}, for which we derive $M_V=-7.6\pm0.3$ and $r_h=460^{+178}_{-90}$~pc, compared to literature values of $M_V=-8.4\pm0.4$ and $r_h=680^{+250}_{-140}$~pc \citep{Martin16}. 

We use these values throughout the remainder of this paper. A full re-analysis of the structural fits, particularly for the faintest galaxies, is the subject of future investigation as part of our Treasury program.

\section{Checking the Robustness of Distances to the M31 Satellites}
\label{sec:sys}
The distances presented in Tab.~\ref{tab:DM}, and the quoted confidence intervals, have been computed to account for our uncertain knowledge of the RR Lyrae metallicity, pulsation parameters, and of the adopted PWZ calibration. We now use a set of internal and external consistency checks as a means of exploring subtle systematic effects that could be the result of data reduction, methodology, and our choice of distance anchors (e.g., our treatment of RRc variables, choices in PWZ).


In this section, all the differences in distance are formulated as {$\Delta \mu = \mu_{PWZ} - \mu_{Comp}$}, with $\mu_{PWZ}$ being the distance reported in Tab.~\ref{tab:DM} and $\mu_{Comp}$ being the alternative distance determination we are comparing against.

\begin{figure*}
	\includegraphics[width=\textwidth]{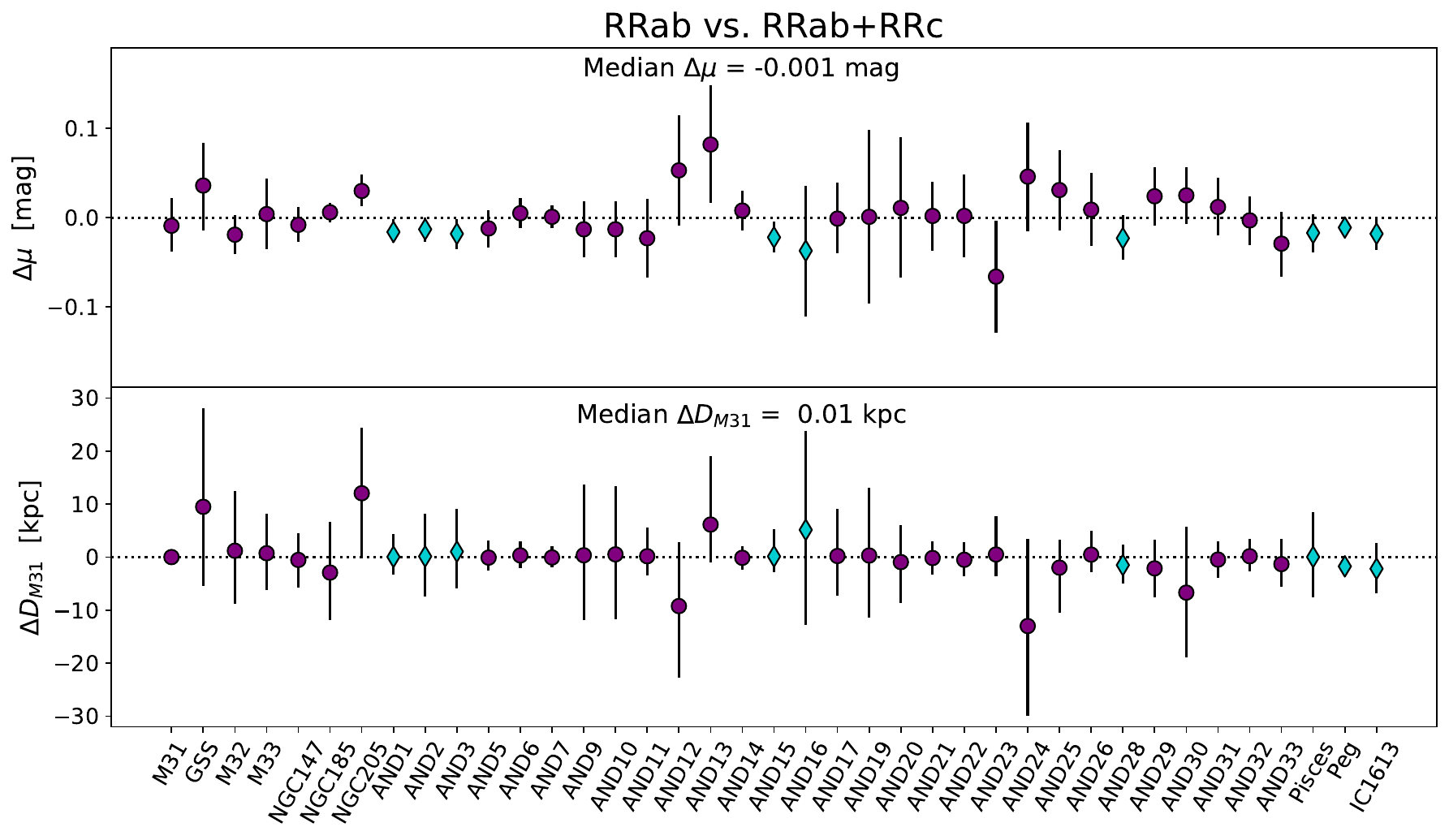} 
	\caption{Difference in the distance modulus (top panel) and in the distance to M31 (bottom panel) resulting from the inclusion of RRc variables in our fits. Purple circles and cyan diamonds represent galaxies with F606W/F814W and F475W/F814W data, respectively.  }
	\label{fig:dRRc}
\end{figure*}
 
 \begin{figure*}
	{\includegraphics[width=\textwidth]{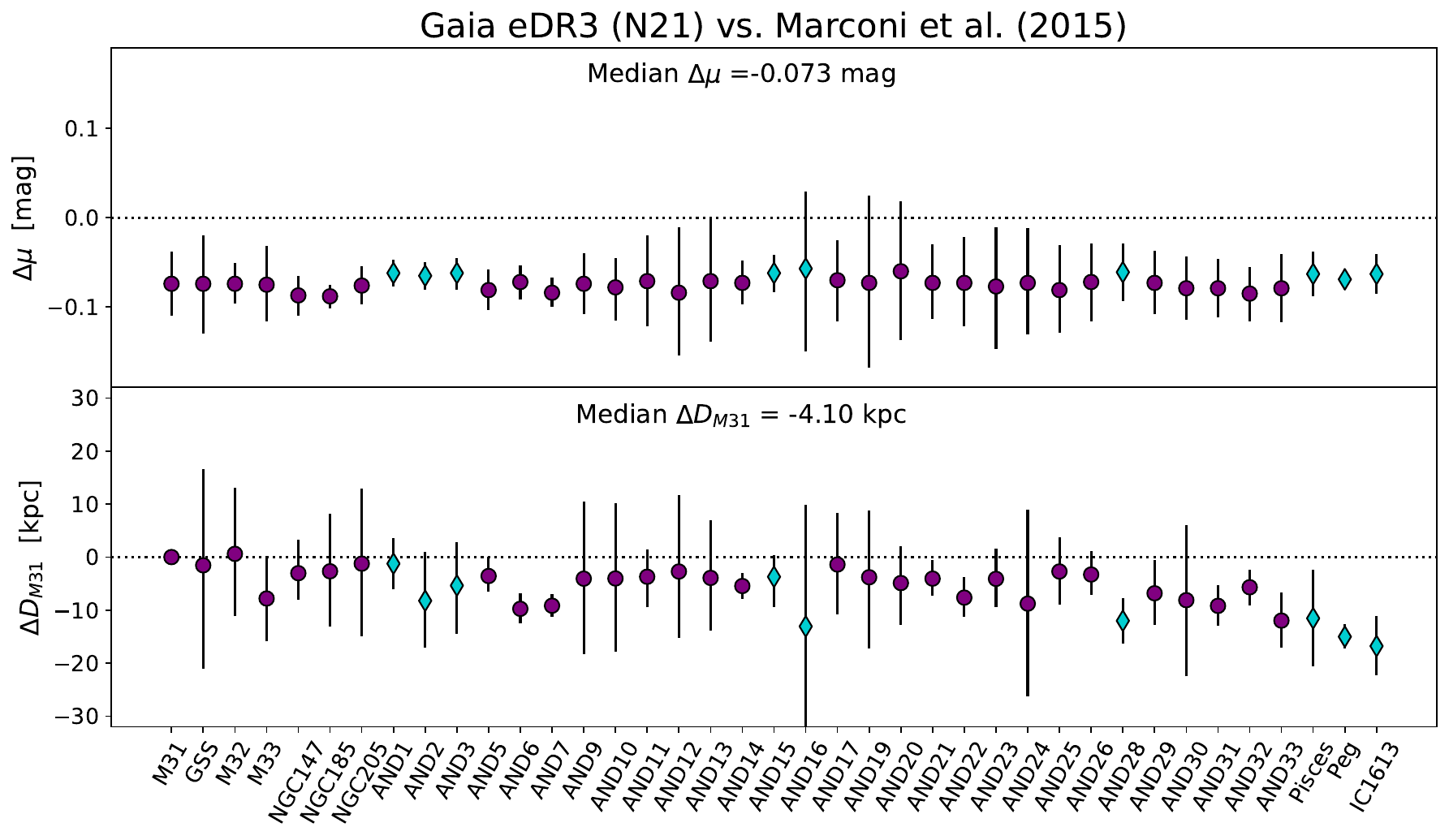}}
\caption{Difference in the distance modulus (top panel) and in the distance to M31 (bottom panel) resulting from the omission of the empirical correction term (Eq.~\ref{eq:Gaia}). Purple circles and cyan diamonds represent galaxies with F606W/F814W and F475W/F814W data, respectively.}
 \label{fig:dEDR3}
\end{figure*}
 
 \subsection{PWZ Distances: the Treatment of RRc Variables}
 \label{sec:RRc}
 Our RR Lyrae sample consists of both RRab and RRc variables. However, the distances of Tab.~\ref{tab:DM} are based on RRab pulsators alone. We made this choice to avoid uncertainties that can arise from the more challenging theoretical modeling of RRc pulsators and to their less robust observational characterization (due to light-curve
shape, shorter periods, and smaller amplitudes). Verifying the impact of this choice and quantifying how the inclusion of RRc pulsation properties affect our distances provides insight into the level of systematics arising from the stellar pulsation models. To explore this, we have run our PWZ models on the full RR Lyrae sample and fit the PWZ of RRc stars using the appropriate coefficients from Tab.~\ref{tab:PWZ}, which are calibrated on the first-overtone pulsation period, $P_{FO}$. We evaluate the correction term of Eq.~\ref{eq:Gaia} using the fundamentalized period \citep[$\log{P_F}=\log{P_{FO}}+0.127$, e.g.,][]{Braga15}.
 
 Figure~\ref{fig:dRRc} shows the difference betwen the distances (heliocentric and relative to M31) determined from RRab only (Tab.~\ref{tab:DM}) and from the full RR Lyrae sample (i.e., RRab and RRc). The inclusion of RRc stars in our fit has a modest effect. Over the whole sample, the distance modulus variation is symmetric around zero, showing no significant systematic bias. The median \textit{absolute} deviation in $\mu$ (top panel) is  0.013~mag (0.6\% change in distance from the MW), which is generally consistent with statistical uncertainties.  The largest effect is observed on AND XIII (0.082~mag; 3.8\%). This galaxy represents one of the lowest signal-to-noise cases in our sample, having both the lowest number of photometric epochs (9 per filter) and a limited number of RR Lyrae (14) used for distance determination. The effect on the relative distances to M31 (bottom panel) is also generally small, with no indication of strong biases. The median \textit{absolute} deviation in relative distance is 0.5~kpc (0.4\% variation). The largest difference is observed for AND XXIV (13~kpc; 6.7\%).
 
In spite of these small differences, we found that the increased sample resulting from the inclusion of the RRc variables does not generally increase our distance precision (due to the dominant contribution of the metallicity term on the error budget). Therefore we made the conservative choice of only measuring distances from the RRab stars and avoid the added uncertainties of first-overtone models.

 \subsection{PWZ Distances: the Effect of the Gaia eDR3 Corrective Term}
 \label{sec:Gaia}
 Another choice in our modeling is the application of the empirical correction term of Eq.~\ref{eq:Gaia}.  The purpose of this is to bring our measurements onto the \textit{Gaia}\ eDR3 distance scale (\hyperlink{cite.Nagarajan22}{N21}) rather than leaving them on the \hyperlink{cite.Marconi15}{M15} distance scale.
 
Figure~\ref{fig:dEDR3} shows differences in the distance moduli and the relative distances to M31 due to the \textit{exclusion} of Eq.~\ref{eq:Gaia} from our framework. The exclusion of the eDR3-based correction term has a non-negligible systematic effect on the distance moduli (top panel). Retaining the \hyperlink{cite.Marconi15}{M15} original distance scale systematically increases the measured distance by a median of 0.073~mag (3.4\% or $\sim 26$~kpc at the distance of M31).  The effect is relatively constant among our sample and the largest variation, observed in NGC~185, is 0.088 mag (4.1\%). This is inline with expectations from Fig.~\ref{fig:Correction}, consistent with the findings of \hyperlink{cite.Nagarajan22}{N21}. We discuss this further in \S~\ref{sec:Improvement}.

Because we self-consistently use the distance to M31 as an anchor point, the relative distances are much less affected by our choice of distance scale (bottom panel). The relative distances to M31 increase by a median amount of 4.1~kpc (3.4\% variation), with a maximum change of 17~kpc in the case of the Pegasus DIG (3.3\% of the 511~kpc distance from M31). As a comparison, the average random uncertainty in the relative distances is $\sim10$~kpc or 8\%.

\subsection{PWZ Distances: the Treatment of the RR Lyrae Metallicity}
\label{sec:RRMet}

Our distance uncertainties are primarily driven by the unknown RR Lyrae metallicity.  It is therefore useful to examine the choices we made for treatment of metallicty and its effect on distances. 

Our first decision is to adopt a single [Fe/H] value for the entire RR Lyrae population. This assumption is not critical, since the RR Lyrae metallicity spreads quantified by observational \citep[e.g., ][]{Clementini05} and theoretical studies \citep[e.g.,][]{Savino20} are a sub-dominant contributor to scatter in the PWZ, compared to the measurement uncertainties on W. To further verify this, we repeated our analysis with a 0.3~dex spread in the assumed RR Lyrae metallicity and found it affected our distances by $<0.01$~mag. \hyperlink{cite.Nagarajan22}{N21} (see Appendix~E) comes to a similar conclusion.

The choice of the metallicity prior has a more significant impact. We could have chosen to adopt a fixed value of [Fe/H] \citep[e.g., ][]{Martinez-Vazquez17,Oakes22}, or to use a tighter prior when sampling the posterior. While this would have lowered our precision floor, it would have increased the contribution of metallicity to the systematic error budget.

For example, it seems intuitive to use a prior based on the [Fe/H] that was informed by RGB star spectroscopy of a given galaxy \citep[e.g., ][]{Collins13} or by the luminosity-metallicity (L-Z) correlation observed in local galaxies \citep[e.g., ][]{Kirby13}. However, such choice is less justified than it might appear at first glance. This is because the characteristic metallicity of the RR Lyrae population is not guaranteed to be representative of the average galaxy metallicity. In fact, due to the specific effective temperature range required to trigger radial pulsation in HB stars, RR Lyrae are produced by a very specific subset of a galaxy's stellar population, whereas RGB star metallicites come from a broader range of populations. As a result, the typical metallicity of the RR Lyrae population is only weakly dependent on the galaxy metallicity distribution function.  

Instead, RR Lyrae metallicities are much more sensitive to a galaxy's SFH. Counterintuitively, RR Lyrae metallicites are higher for older stellar populations.  This concept was first shown by \citet{Lee92}, and further quantified by \citet{Savino20}.  More concretely, Fig.~\hyperlink{https://www.aanda.org/articles/aa/pdf/2020/09/aa38305-20.pdf}{10} of \citet{Savino20} clearly shows how the metallicity of the RR Lyrae can differ by as much as 1.5~dex from the average RGB metallicity.

This effect is unlikely to be severe for dwarfs of intermediate luminosity, for which the L-Z relation predicts average metallicities close to what inferred through the models of \citealt{Savino20}.  However, it can have a significant impact on the brightest (faintest) galaxies.  In these regimes, the RR Lyrae production efficiency is expected to peak at much lower (higher) metallicities than the average RGB metallicity. We quantify this effect by re-running our models and swapping our flat [Fe/H] prior for a Gaussian prior, with standard deviation of 0.5~dex, and centered on the [Fe/H] predicted by the L-Z relation of \citet{Kirby13}. Over the whole sample, the median \textit{absolute} change in distance modulus, compared to the values of Tab.~\ref{tab:DM}, is 0.06 mag or $\sim 3\%$ in heliocentric distance. For galaxies in the luminosity range $-16.5<M_V<-7$ ($\sim 70\%$ of the sample), the inferred distances change by less than 0.1~mag from the values of Tab.~\ref{tab:DM}, while for galaxies outside this interval the change is larger. The strongest effect is observed in M31 itself, for which a prior centered on $\rm[Fe/H] = -0.4$ \citep[consistent with RGB metallicities in the inner halo, e.g., ][]{Gilbert14}, results in a distance of $\sim 716$~kpc ($\Delta \mu =0.18$ mag), much lower than the canonically accepted range of 760-780~kpc (\S~\ref{sec:M31}). This discrepancy is understood by considering that the metal rich helium-burning stars in the M31 halo mostly occupy the red clump and therefore do not contribute to the RR Lyrae population, which is generated by more metal-poor stars. A prior of $\rm[Fe/H] = -0.4$ for the RR Lyrae is therefore not well motivated.

The argument for M31 applies, to lesser degree, to all galaxies for which the L-Z based metallicity results in an inefficient RR Lyrae production. Given these difficulties in adopting a physically consistent prior on [Fe/H], we have chosen, in \S~\ref{sec:Model} to be agnostic about the RR Lyrae metallicity, except for the boundaries of the prior, which were chosen to limit [Fe/H] to values that are expected to produce RR Lyrae efficiently.

\begin{figure}

\plotone{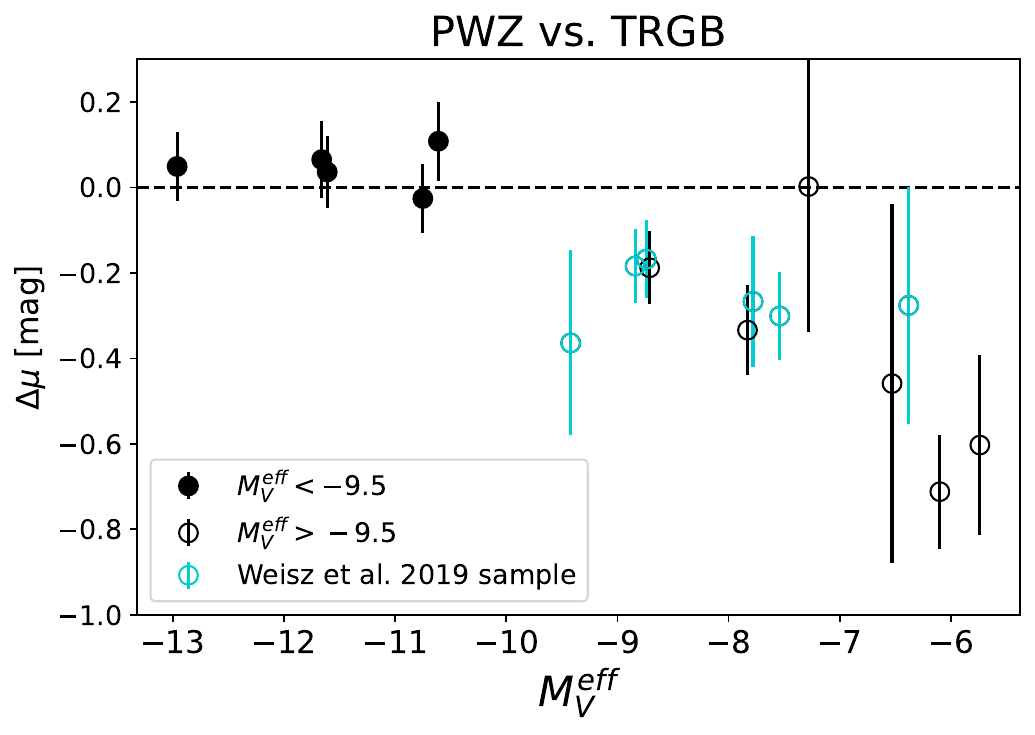}
\caption{Difference between the distance modulus obtained through our PWZ modeling and that obtained by fitting the TRGB magnitude. Filled symbols represent galaxies with $M^{\rm eff}_V<-9.5$, which are part of our validation sample. Empty symbols are galaxies with $M^{\rm eff}_V>-9.5$. Galaxies highlighted in cyan are part of the TRGB anchor sample of \citet{Weisz19}. RR Lyrae are necessary for accurate and precise distances to faint galaxies in and around the Local Group (e.g., those that Rubin Observatory should discover).}
\label{fig:TRGB}
\end{figure}

\subsection{Comparison with the Tip of the Red Giant Branch}
\label{sec:TRGB}

The TRGB is another well-established Pop II distance indicator \citep[see ][and references therein]{Beaton18}.  Due to its brightness, the TRGB can be used to much larger distances than RR Lyrae, making it excellent for mapping the Local Volume and anchoring $H_0$ \citep[e.g.,][]{Tully16,McQuinn17, Freedman20}.  However, unlike RR Lyrae, a large number of stars are necessary to clearly identify the TRGB and measure precise and/or accurate distances \citep{Madore95}. In this section, we measure TRGB distances to select galaxies in our sample to better quantify the consistencies and limitations of these Pop II distance indicators.

We selected a TRGB validation sample comprising of all the galaxies that, in our ACS field of view, have a sufficiently high number of stars to allow a robust measurement. We quantified this number by using total absolute magnitudes, sizes, ellipticities, and position angles of each galaxy (from Tab.~\ref{tab:DM} and references therein) to build light-profile models for our target galaxies. We used exponential light profiles for all galaxies except M32, NGC~147, NGC~185, and NGC~205, for which we used S\'ersic profiles (indexes provided in the caption of Tab.~\ref{tab:DM}).  We calculated the fraction of the total galactic light captured in the ACS field of view and derived an effective absolute magnitude, $M^{\rm eff}_V$, for our photometric catalogs (e.g., an ACS field that contains 10\% of the galaxy light would result in a $M^{\rm eff}_V$ 2.5 magnitudes fainter than the galaxy's $M_V$). We selected galaxies that have $M^{\rm eff}_V<-9.5$. This corresponds to a stellar mass of roughly $10^6 M_{\odot}$, comparable to globular clusters such as $\omega$ Cen and 47 Tuc.  This is generally regarded as the lower limit for reliable TRGB measurements \citep[e.g.,][]{Bellazzini01,Bellazzini04,Bono08,Soltis21}. We excluded M32, as its projected proximity to the disc of M31 could introduce significant uncertainties in the amount of foreground extinction and contamination from M31's red giant population. We also excluded galaxies with F475W/F814W data, as there is no TRGB calibration for that filter combination in the literature. Our validation sample is therefore composed by: NGC~147, NGC~185, NGC~205, And~{\sc VI} and And~{\sc VII}. While these galaxies are the only ones that can provide a meaningful comparison with our RR Lyrae-based distances, we also attempted to measure the TRGB in galaxies with fainter $M^{\rm eff}_V$, to explore the robustness of TRGB measurements as a function of observed stellar population size.

We measured the observed TRGB magnitude using the procedure described in \citet{McQuinn16a,McQuinn16b}, which uses the TRGB calibration of \citet{Rizzi07}. We dereddened the photometry using the extinction map of \citet{Schlafly11}. Then we identified the observed TRGB magnitude using a maximum likelihood modeling of the RGB luminosity function that takes into account photometric uncertainties and completeness.  We define the uncertainty in the TRGB distance as the quadrature sum of the uncertainty in the apparent TRGB magnitude, the uncertainty on the \citet{Rizzi07} calibration \citep[as prescribed in][]{McQuinn17}, and the uncertainty in the dust correction. We took the dust extinction uncertainty as 10\% of the \citet{Schlafly11} extinction \citep[as motivated in the original calibration of][]{Schlegel98} with a floor of 0.02 in E(B-V) \citep[to account for the systematic uncertainties reported in the re-calibration of][]{Schlafly11}.

Figure~\ref{fig:TRGB} presents a comparison between our RR Lyrae-based distances and our TRGB distances. For all the galaxies in the validation sample (filled dots), the two distances agree within 0.1 mag and are well-within the measurement uncertainties. In this luminosity regime, we found that TRGB distance moduli are systematically smaller than RR Lyrae distance moduli, with a median difference of 0.049 mag (2.3\% distance difference). Again, this effect is comparable with the random uncertainties of the PWZ distances.  

For the 16 galaxies with $M^{\rm eff}_V>-9.5$ for which the TRGB modeling converged (empty symbols), the TRGB distance is systematically larger than inferred through the RR Lyrae, with the difference strongly increasing with fainter $M^{\rm eff}_V$ (up to 0.7 mag, or $\sim 35\%$ distance difference, for $M^{\rm eff}_V \approx -6$), i.e., the magnitude of TRGB is measured to be too faint with respect to what predicted from the RR Lyrae distance. This trend is a result of the stochastic RGB sampling in low-luminosity galaxies and it is a clear demonstration of the need of a large stellar sample for robust TRGB measurements, as first demonstrated by \citet{Madore95} and further discussed by, e.g., \citet{Makarov06,Conn11,McQuinn13}. 

\setcitestyle{notesep={;}}

The comparison of Fig.~\ref{fig:TRGB} provides an empirical quantification of the luminosity regime below which measured TRGB magnitudes lose fidelity as a distance indicator. In principle, well calibrated HB magnitudes could provide a robust distance scale that extends below this luminosity threshold. However, the HB suffers from similar stochasticity problems as the TRGB and reliable HB magnitudes become hard to measure for galaxies with $M_V\gtrsim -7$. RR Lyrae stars, on the other hand, can provide robust distances even in a small-sample regime, and have been identified in stellar systems as faint as $M_V=-2.7$ \citep{Martinez-Vazquez19}. Therefore, outside the MW halo, they remain the sole option to attain a precise distance scale that is consistent over a large dynamic range of luminosities and SFHs. This is a particularly important takeaway in view of the many faint stellar systems that should be uncovered by wide-area searches for galaxies in and beyond the Local Group \citep[e.g., ][ searches of Roman/Rubin data]{Simon19,Sand22}.

\setcitestyle{notesep={,}}

\begin{figure*}

\plotone{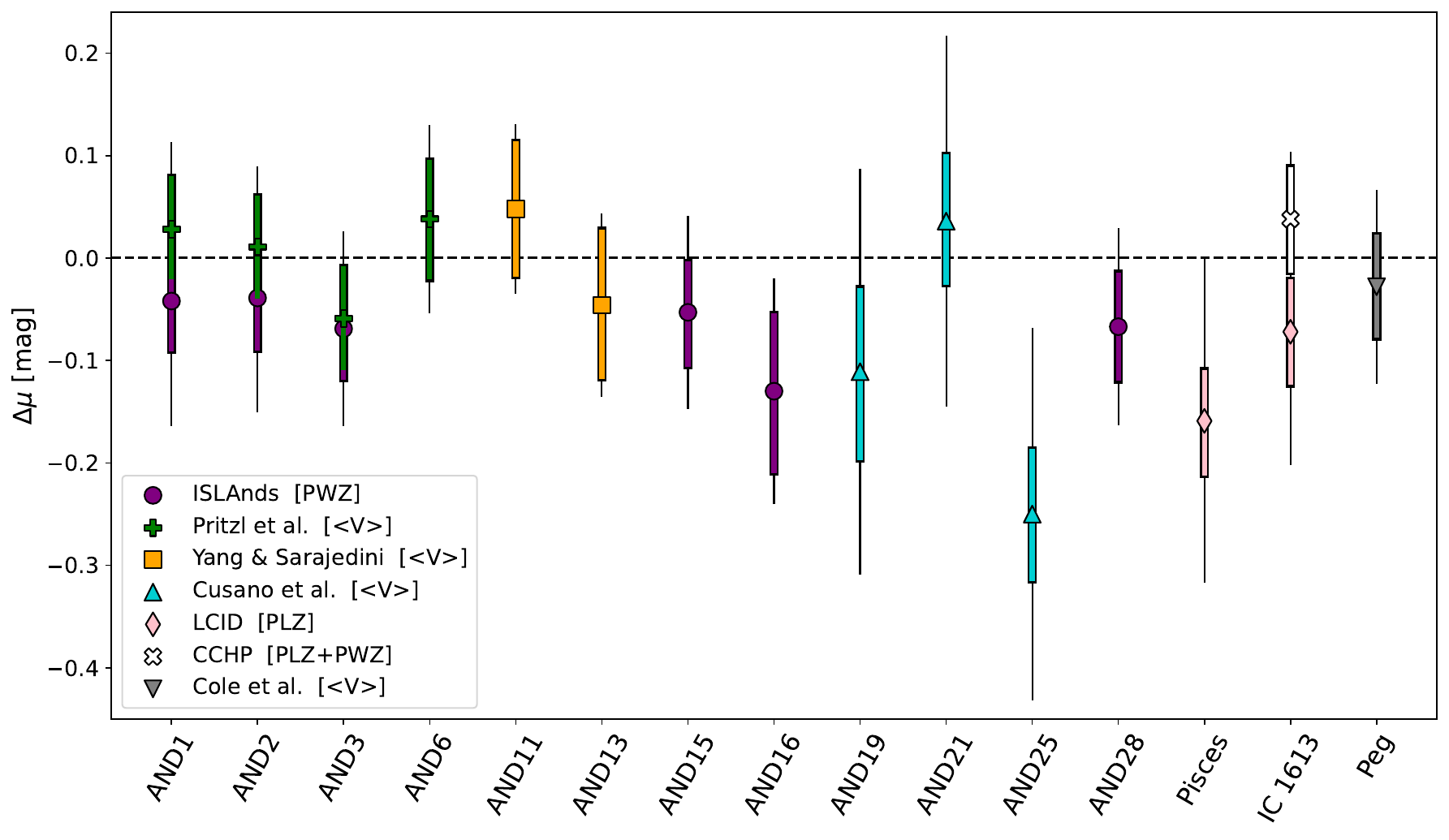}
\caption{Difference between the distance modulus calculated in this paper and literature values from other RR Lyrae-based analyses \citep{Pritzl02,Pritzl04,Pritzl05,Bernard10,Hidalgo11,Yang12,Cusano13,Cusano15,Cusano16,Cole17,Hatt17,Martinez-Vazquez17}. The thick colored portion of the error bar shows the contribution to the total error budget coming from our distance measurements.}
\label{fig:Lit_RRL}
\end{figure*}

\begin{figure*}

\plotone{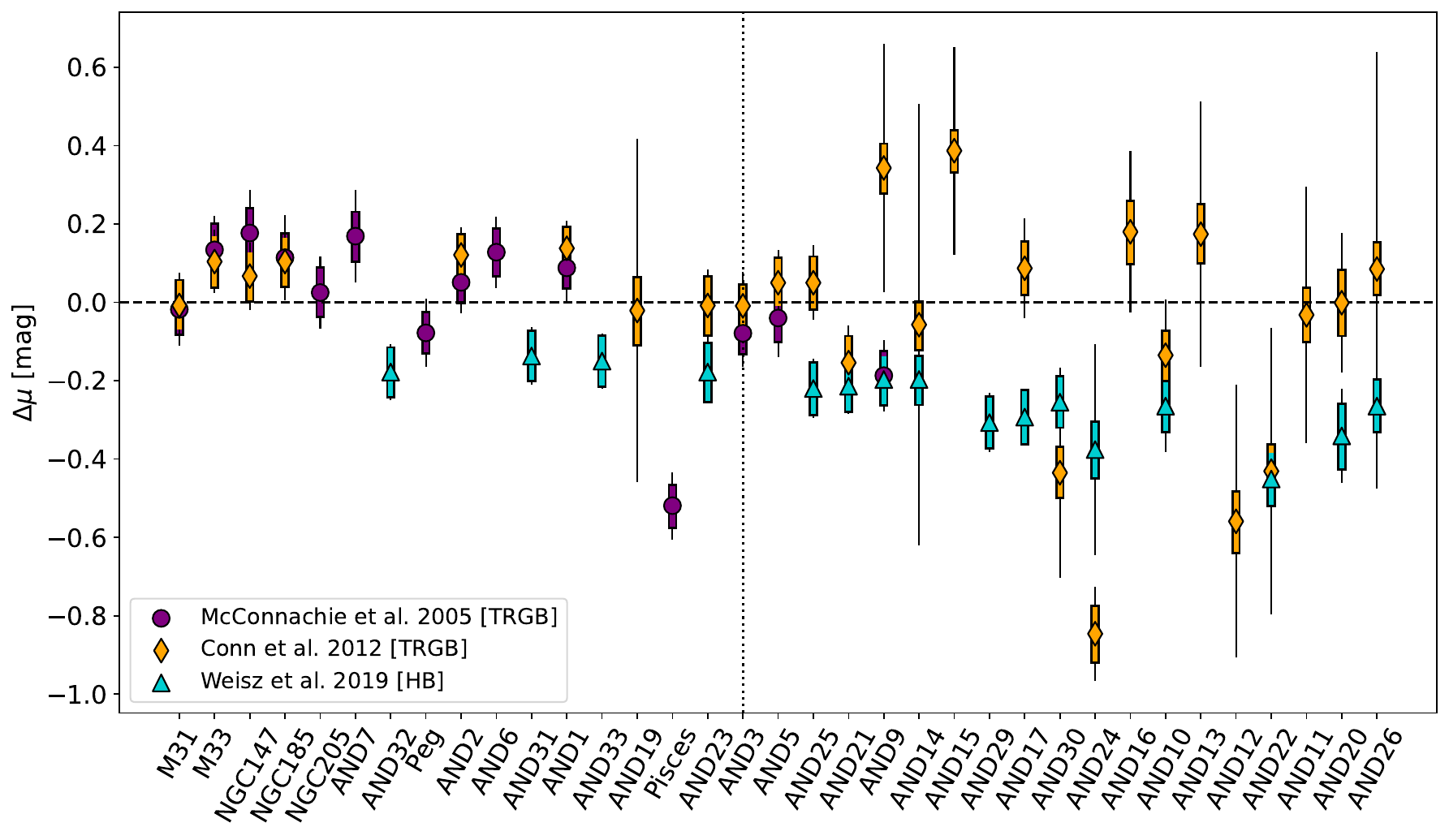}
\caption{Difference between the distance modulus calculated in this paper and literature values from the compilations of \citet[purple circles]{McConnachie05}; \citet[orange diamonds]{Conn12} and \citet[cyan triangles]{Weisz19}. The thick colored portion of the error bar shows the contribution to the total error budget coming from our distance measurements. The galaxies are ordered in descending order of absolute luminosity. The vertical dotted line marks $M_V = -9.5$.}
\label{fig:Lit_Misc}
\end{figure*}

\pagebreak
\subsection{Comparison with Previous RR Lyrae Distances}
\label{sec:RRLit}
A subset of M31 satellites have RR Lyrae distances in the literature.  These are generally derived for single galaxies or small sets using a range of observations and techniques. The largest set of M31 satellite RR Lyrae distances (6 galaxies; And~{\sc I}, And~{\sc II}, And~{\sc III}, And~{\sc XV}, And~{\sc XVI}, and  And~{\sc XXVIII}) were published by \citet{Martinez-Vazquez17} as part of the \hst\ ISLAndS program \citep{Skillman17}. This study derived the variable star properties using the same \hst/ACS observations of this paper and obtained distances using PWZ relations based on \hyperlink{cite.Marconi15}{M15} models. However, their results are based on different photometric reductions, light-curve analyses, and PWZ formalism. The distance to the Pisces and IC~1613 dwarf galaxies \citep{Bernard10,Hidalgo11} were studied as part of the LCID program \citep{Gallart15}. The observations used are also the same of this paper (although in Pisces we complement the F814W data with additional observations from GO-13738), but the distances were derived using a period-luminosity-metallicity (PLZ) relation. The RR Lyrae of IC~1613 were also studied in the context of the Carnegie-Chicago Hubble Program \citep[CCHP,][]{Beaton16}, and a distance was derived using a combination of PWZ and PLZ from a completely independent \hst\ dataset \citep{Hatt17}.  Finally the mean V magnitude of the RR Lyrae population has been used as a distance proxy for And~{\sc XIX}, And~{\sc XXI}, and And~{\sc XXV} using ground-based data \citep{Cusano13,Cusano15,Cusano16}; for And~{\sc I}, And~{\sc II}, And~{\sc III}, And~{\sc VI}, And~{\sc XI} and And~{\sc XIII} using \hst/WFPC2 observations \citep{Pritzl02,Pritzl04,Pritzl05,Yang12}; and for the Peg~DIG using \hst/ACS data \citep{Cole17}.

Figure~\ref{fig:Lit_RRL} shows differences in RR Lyrae distances from this paper and the literature. The distances of the largest sample, that of \citet{Martinez-Vazquez17}, are systematically larger by a median of 0.06~mag. This effect is consistent with the eDR3 correction (Eq~\ref{eq:Gaia}), which is known to produce distances that are $\sim$0.06~mag closer than \hyperlink{cite.Marconi15}{M15} (\hyperlink{cite.Nagarajan21}{N21}).

The comparison with other studies shows a somewhat larger scatter, which may reflect the variety of techniques used. We find a median $\Delta \mu=-0.028$~mag for the ensemble. On an individual galaxy basis, most differences are at the level of 0.1~mag and are consistent within the reported uncertainties. The only notable exception to this trend is And~{\sc XXV}, for which the existing literature distance is 0.25~mag larger than that determined herein. The cause of this discrepancy is not clear, but existing literature measurements for this galaxy show a similar degree of disagreement (e.g., \citealt{Conn12} measured a distance to And~{\sc XXV} of $\mu=24.33^{+0.07}_{-0.21}$, as opposed to the value of $\mu=24.63\pm0.17$ derived by \citealt{Cusano16}). A detailed exploration of this discrepancy is not within the scope of this paper. Overall, the scatter observed in Fig.~\ref{fig:Lit_RRL} exemplifies the effect of methodological differences on available literature distances, and highlights the importance of large self-consistent compilations. Such lack of homogeneity in the literature was recently highlighted by the review of \citet{Monelli22}, who estimate systematic effects that are in good quantitative agreement with the comparison shown here.

\begin{figure*}

\subfloat[][]
 	{\includegraphics[width=\textwidth]{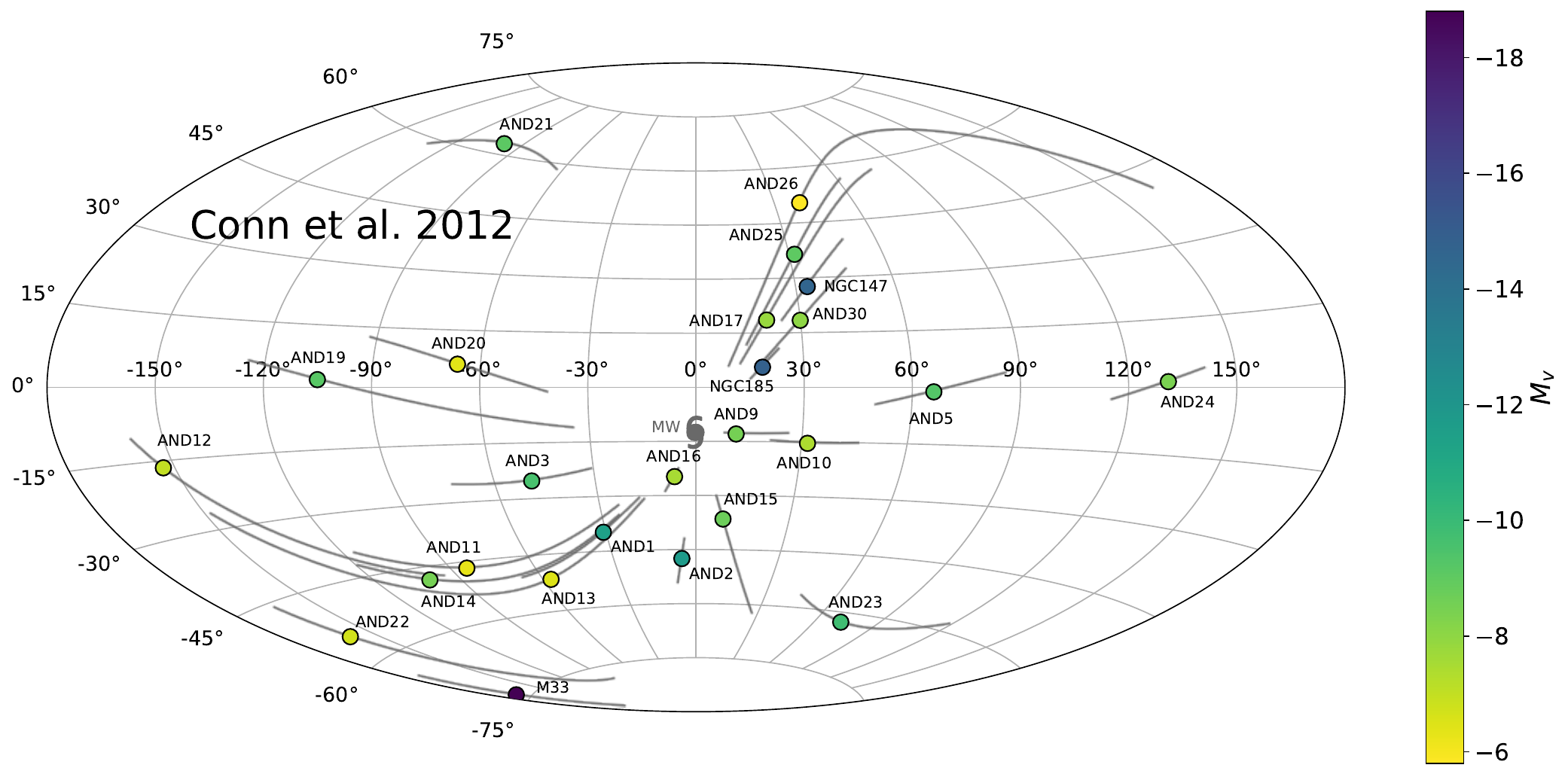} \label{fig:SkyC12}} \quad
\subfloat[][]
	{\includegraphics[width=\textwidth]{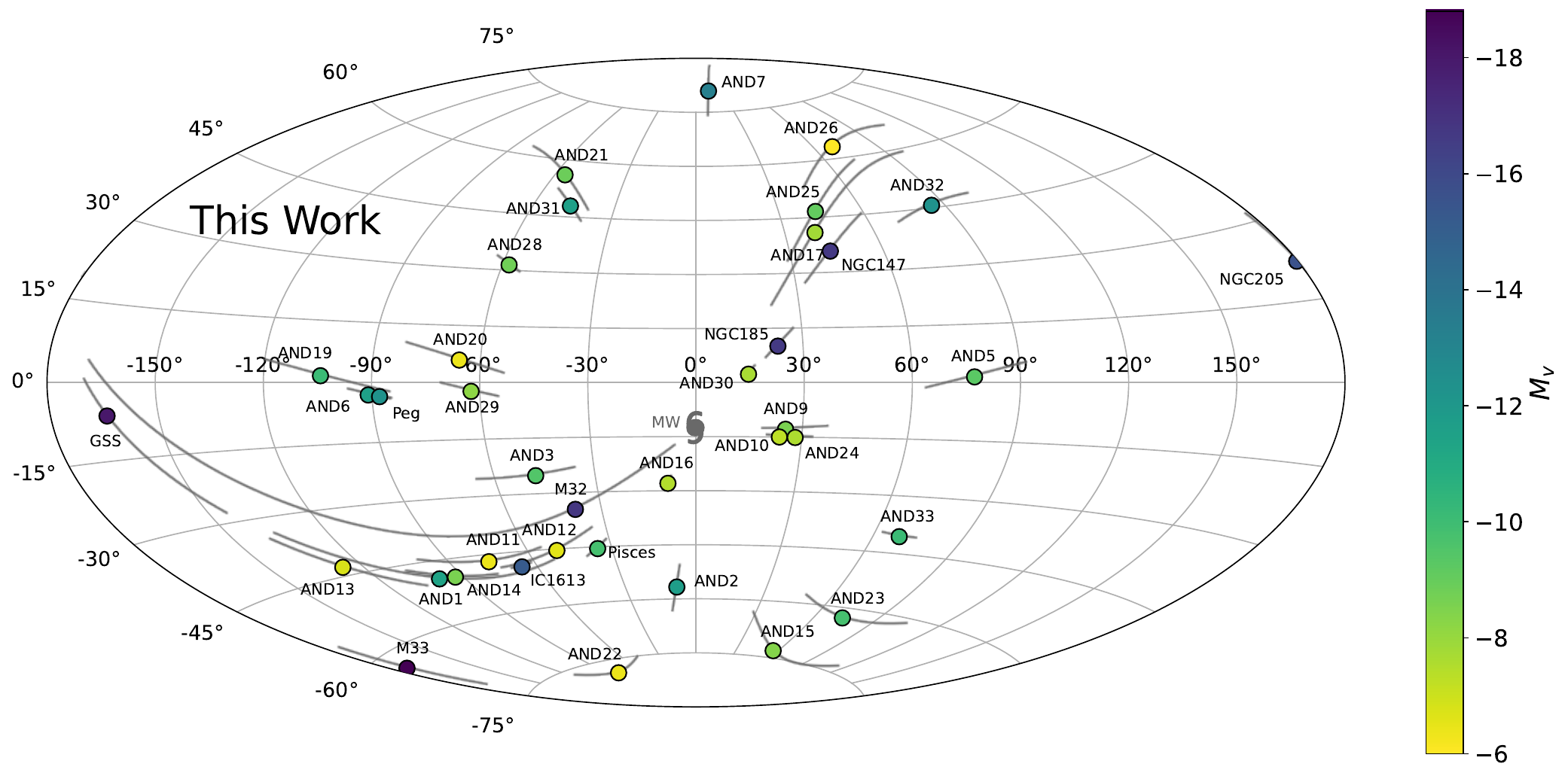} \label{fig:SkyNew}} \quad

\caption{a) Sky projection (in the galactic coordinate system of \citealt{Conn12}) of the M31 satellites, color coded by absolute luminosity, as seen from the center of M31, using the distance data of \citet{Conn12}. The gray lines represent the uncertainties in the sky position. The direction of the MW is also reported. b) Same as Fig.~\ref{fig:SkyC12}, but using the distances calculated in this paper. The comparison illustrates the improvement in sample size and distance precision of our compilation. The few remaining galaxies with large positional uncertainties are those that lie in close proximity to M31 ($D_{31}\lesssim60~$kpc).}
\label{fig:Sky}
\end{figure*}

\subsection{Comparison with Other Literature Compilations}
\label{sec:MiscLit}

The distances presented in this paper represent the first homogeneous analysis of nearly the full M31 system as we know it today.  However, there are already a few notable compilations that comprise a significant sample of M31 satellites. Using TRGB measurements from ground-based observations, \citet{McConnachie05} derived distances for 14 M31 satellites and M31 itself. The TRGB from ground-based observations was also used by \citet[][hereafter C12]{Conn12} on 26 M31 satellites, extending the TRGB measurements to a number of fainter galaxies discovered by the PAndAS survey \citep{Martin06,Ibata07,McConnachie08,McConnachie09,Martin09,Richardson11,McConnachie18}, while \citet{Weisz19} derived distances to 16 galaxies using the observed magnitude of their HB.

Figure~\ref{fig:Lit_Misc} presents a comparison between our RR Lyrae-based distances and the results from these three compilations. Galaxies are plotted in order of decreasing luminosity. To guide the reader, we have indicated with a vertical line $M_V = -9.5$, which is the luminosity limit we have used to define our TRGB validation sample (\S~\ref{sec:TRGB}). Visually, this plot reveals significant scatter among the literature compilations themselves. For the same galaxy, differences on the order of 0.2 mag are not uncommon and, in some cases, the distances differ by up to 0.5 mag. In the following sections, we discuss the trends related to this figure in more detail.

\subsubsection{Literature TRGB Distances}

For galaxies brighter than And~{\sc XXV} ($M_V = -9.1$), we find reasonable consistency between literature TRGB determinations and our RR Lyrae distances, although our distances are on average larger than both those of \citet{McConnachie05} and \hyperlink{cite.Conn12}{C12}. For galaxies with $M_V<-9.5$, the median difference is 0.067 mag, only slightly larger than what we quantified in \S~\ref{sec:TRGB}, in spite of \citet{McConnachie05} and \hyperlink{cite.Conn12}{C12} using different TRGB calibrations and dust corrections.

For the faint galaxies, the comparison with the TRGB distances shows a large scatter. This is not surprising, as we have already shown how TRGB measurements in this luminosity regime can be heavily affected by stochastic sampling, and it is consistent with the size of the TRGB-based distance uncertainties, which rapidly increase for galaxies with $M_V \gtrsim -9.5$. We do not observe, however, a defined trend as a function of galaxy magnitude, possibly owing to the different technique that \hyperlink{cite.Conn12}{C12} uses to measure the TRGB.

\subsubsection{Literature HB Distances}
The HB distances, on the other hand, are overall 0.25 mag (median) larger than our PWZ-based distances. Ultimately, the calibration of the \citet{Weisz19} HB distance scale is anchored on TRGB measurements of eight galaxies in their sample. For six of these eight galaxies (And~{\sc IX}, And~{\sc XX}, And~{\sc XXIX}, And~{\sc XXXI}, And~{\sc XXXII} and And~{\sc XXXIII}), we successfully measured the TRGB distance in \S~\ref{sec:TRGB} (cyan symbols in Fig.~\ref{fig:TRGB}). All of these galaxies have overestimated TRGB distances due to sparsely populated RGBs, with a median difference of 0.27 mag compared to the PWZ-distances. As \citet{Weisz19} use similar \hst/ACS observations and measure the TRGB using our same technique and calibration, we expect a systematic of similar size to be present in their measurements. This effect is then propagated to the HB distance scale.  This comparison serves to illustrate the importance of fully understanding systematic effects on extragalactic distance anchors as well as limitations of the TRGB for faint galaxies.

\subsection{A Larger, Homogeneous Distance Catalog to the M31 System: Quantifying Uncertainties}
\label{sec:Improvement}


The extensive set of internal and external comparisons  presented in this section reveal the high accuracy of our distance measurements. Changes to our distance anchor (\S~\ref{sec:RRc}, \ref{sec:Gaia}) and cross-checks with the TRGB (\S~\ref{sec:TRGB}) resulted in a maximum systematic variation of 3.4\% in the distances. This corresponds to a difference $\sim 26$~kpc (0.07 mag) from the MW and of $\sim 4$~kpc from M31. Similarly, axamination of our metallicity assumptions (\S~\ref{sec:RRMet}) and comparisons with literature measurements (\S~\ref{sec:RRLit}, \ref{sec:MiscLit}) indicate that systematic differences are generally contained within 0.1 mag.

Compared to previous compilations, our RR Lyrae distances are a significant improvement in homogeneity and precision, enabling more robust characterizations of the M31 satellite system global properties. The largest literature M31 distance catalog, prior to this work, was that of \hyperlink{cite.Conn12}{C12}, which represented a cornerstone for determinations of the spatial and structural properties of the M31 system \citep[e.g.,][]{Conn13, Geraint13,Watkins13,Martin16,Salomon16,McConnachie18}. Compared to \hyperlink{cite.Conn12}{C12}, our work increases the galaxy sample by $\sim$ 50\% and by using RR Lyrae as a distance indicator, we improve on distance precision at all galaxy luminosities.
The median precision of our distances is 21~kpc (2.9\%). Our largest distance uncertainty is 32~kpc (3.9\%; And~{\sc XIX}).  These are substantially lower than the median error of \hyperlink{cite.Conn12}{C12} (36~kpc) and their largest uncertainty for faint galaxies which is 190~kpc (And~{\sc XXVI}).

Following \hyperlink{cite.Conn12}{C12}, Figure~\ref{fig:Sky} shows a sky-projected view of the  stellar systems in our sample as it would be observed from the center of M31, comparing the previous map from \hyperlink{cite.Conn12}{C12} (Fig.~\ref{fig:SkyC12}) and our new data (Fig.~\ref{fig:SkyNew}). The sky positions are expressed in galactocentric coordinates, with b=$90^{\circ}$ pointing at M31's north galactic pole and l=$0^{\circ}$ in the direction of the MW.

This figure nicely illustrates the clear improvement that our new data set brings in terms of sample size and sky localization. Many galaxies can now be located in M31's sky within a few degrees and only some of the satellites in the innermost halo (e.g., M32 not present in the \hyperlink{cite.Conn12}{C12} compilation) retain an uncertain sky localization. This is due to their small distance to M31, which translates small errors in distance into large errors on the polar coordinates. In the following section, we leverage our improved distances to examine select structural and spatial properties of the M31 satellite population.

\section{Discussion}
\label{sec:applications}

\subsection{Large Scale Asymmetries in the Satellite Distribution}
\label{sec:Cartesian}

\begin{figure*}

\includegraphics[width=0.95\textwidth]{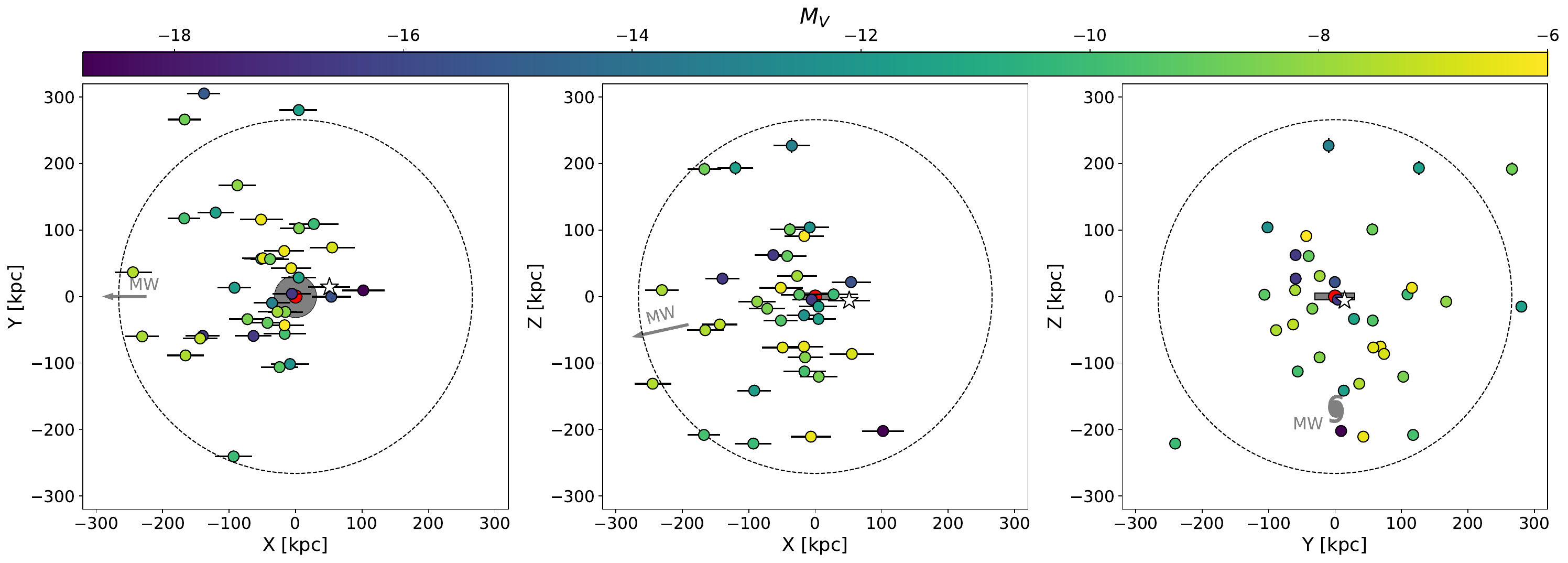}
\caption{Projected views of the M31 system onto the three major planes of the cartesian coordinate system described in \S~\ref{sec:Cartesian}. In this ref-
erence frame, the XY-plane lies onto the M31 disc, with the
positive Z-axis pointing M31’s north galactic pole and the
positive X-axis having the same azimuth, in the M31 disc, as the M31-Earth vector. Galaxies are color-coded by absolute luminosity. The position of our GSS line-of-sight is marked as a white star. The M31 disc (not to scale) orientation and the position of the MW are also reported. The dashed circle traces the virial radius of M31 \citep[266~kpc,][]{Fardal13,Putman21}.}
\label{fig:XYZ}
\end{figure*}
One application of our precise RR Lyrae distances is precise 3D cartography of the M31 satellite system. To do this, we sample the MCMC chains of each dwarf galaxy (from \S~\ref{sec:Model}), in parallel with the chains for M31, to combine the distance posterior probabilities and obtain a probability distribution for the coordinates of the satellite in a Cartesian reference frame centered on M31. For ease of comparison with past work, we orientated the reference frame to be identical to that of \citet{McConnachie06} and \hyperlink{cite.Conn12}{C12}. In this reference frame, the XY-plane lies onto the M31 disc, with the positive Z-axis pointing M31's north galactic pole and the positive X-axis having the same azimuth, in the M31 disc, as the M31-Earth vector.

Figure~\ref{fig:XYZ} presents the projection of the satellite positions onto the three orthogonal Cartesian planes. The most striking feature that can be appreciated from this view is the stark asymmetry in the satellite distribution around M31. Approximately 80\% of the dwarf galaxies around M31 are located in one hemisphere of the M31 halo. This hemisphere also appears to be roughly aligned with the direction of the MW.

The asymmetry in M31 satellites was previously identified by \citet{McConnachie06}, and later by \citet{Conn13}, using TRGB distances, but it is even more pronounced with our new updated distances, as a handful of dwarfs previously beyond M31 are now on its nearside (e.g., And~{\sc XII}, And~{\sc XXII}, and And~{\sc XXIV}). The reason for this asymmetric distribution in M31 is not known.

The alignment of this asymmetry with the MW direction raises the possibility that observational effects could play a role. Recently, \citet{Doliva-Dolinsky22} calculated detailed dwarf galaxy detection limits for the PAndAS survey, which is responsible for finding most faint M31 satellites. They find that the limiting magnitude between the near and far side of M31's virial radius should change by no more than 1 magnitude. This means that, apart from a small sky region directly behind M31's disc, galaxies brighter than $M_V\lesssim -7.5$ should be detectable throughout M31's halo. The persistence of strong asymmetry (at the same $\sim 80\%$ level) among galaxies above this brightness indicates that, if observational effects are playing a role in the observed dwarf galaxy distribution, they are more nuanced than photometric completeness alone. A more detailed model of the spatial properties of M31 satellites, including a quantitative treatment of the spatial completeness maps, is already ongoing (Doliva-Dolinsky, in prep.). 

In the past few years there has been some observational \citep[e.g., ][]{Libeskind16, Brainerd20} and theoretical \citep[e.g., ][]{Pawlowski17,Forero-Romero18,Wang20,Wang21} evidence that at least some lopsidedness can occur in the satellite system of massive hosts, especially if a nearby massive companion (such as the MW) is present. Even so, the M31 system seems to represent a rather extreme configuration. On the other hand, M31 is essentially the only system, apart from the MW, where an accurate 3D reconstruction of the satellite-galaxy system can be obtained, while for more distant galaxy groups only projected spatial distributions can be examined. In the original discovery paper, \citet{McConnachie06} provide a detailed discussion of possible dynamical and cosmological explanations for the M31 lopsided satellite distribution; we refer the reader to this paper for more details. Among several of these explanations, one is that much of the satellite population was accreted recently and have not had time to fully phase mix.  This scenario is broadly consistent with various pictures of M31's recent accretion history \citep[e.g., ][]{Ferguson16,DSouza18,McConnachie18,Mackey19} and could reconcile the extreme properties of M31 with respect to the broader galaxy population. Clearly, further work is needed to clarify the (lack of) consistency of the M31 dwarf galaxy system with the properties of observed and simulated satellite populations \citep[e.g., ][]{Pawlowski21,Boylan-Kolchin22}.

\subsection{The Great Plane of Andromeda Satellites}
\label{sec:plane}

Another large-scale feature that has sparked discussion and controversy in recent years \citep[see, e.g., ][]{Pawlowski21,Boylan-Kolchin22} is a degree of planar alignment among $\sim 50$\% of the dwarf satellites around Andromeda, emerging from both spatial \citep{Koch06,McConnachie06,Conn13} and kinematic \citep{Ibata13,Sohn20} data. While the lack of 6D phase-space information prevents us from firmly establishing whether the ``great plane of Andromeda" (GPoA) is a long-lived feature or a transitory configuration, the reports of similar features in other local galaxies \citep[e.g.,][]{Pawlowski12,Muller17,Muller18} have prompted a debate on the significance of these planar structures for our current understanding of structure assembly \citep[e.g.,][]{Kroupa05,Zentner05,Libeskind05,Sawala16,Fernando17, Fernando18,Pawlowski18,Samuel21,Sawala22}.

We leveraged our larger sample of precise, homogeneous distances to reassess the properties of the GPoA. We modeled the 3D positions of our galaxy sample using a Gaussian Mixture Model, analogous to that described in \S~\ref{sec:distances}. For this model, we excluded IC~1613 and the Peg DIG, because their large distance from M31 could have a high weight on the recovered orientation of the GPoA, but also decreases the likeliness of physical association with the rest of the satellite system. We also excluded the GSS, since a meaningful modeling of this system in the context of the GPoA would require knowledge of the orbital plane within the GSS itself, which our single line of sight does not provide. For the remaining galaxy sample, we followed the approach of \citet{Conn13} and used the Cartesian coordinates of Fig.~\ref{fig:XYZ} to derive the distance of each satellite from a given plane, passing through M31, as:
\begin{equation}
    D_{Plane}=ax+by+cz,
\end{equation}
where $a$, $b$ and $c$ are the orthonormal components of the plane perpendicular vector. We assumed the plane to have a Gaussian density profile along its perpendicular axis, so that $D_{Plane}$ can be described as drawn from a population $\mathcal{N}(0,\sqrt{rms^2_{Plane}+\sigma^2_{DP}})$, with $\sigma_{DP}$ being the measurement uncertainty on $D_{Plane}$ and $rms_{Plane}$ being the intrinsic dispersion of the planar distribution. We assumed a second, spherical isotropic population to be superimposed to the planar component. Finally, we assumed that the radial distance $D_{31}$ of dwarfs that belong to this component is drawn from the distribution $\mathcal{N}(0,\sqrt{rms^2_{Sphere}+\sigma^2_{D31}})$, with $\sigma_{D31}$ and $rms_{Sphere}$ being the observational error and intrinsic dispersion of $D31$.

\begin{figure*}

\plotone{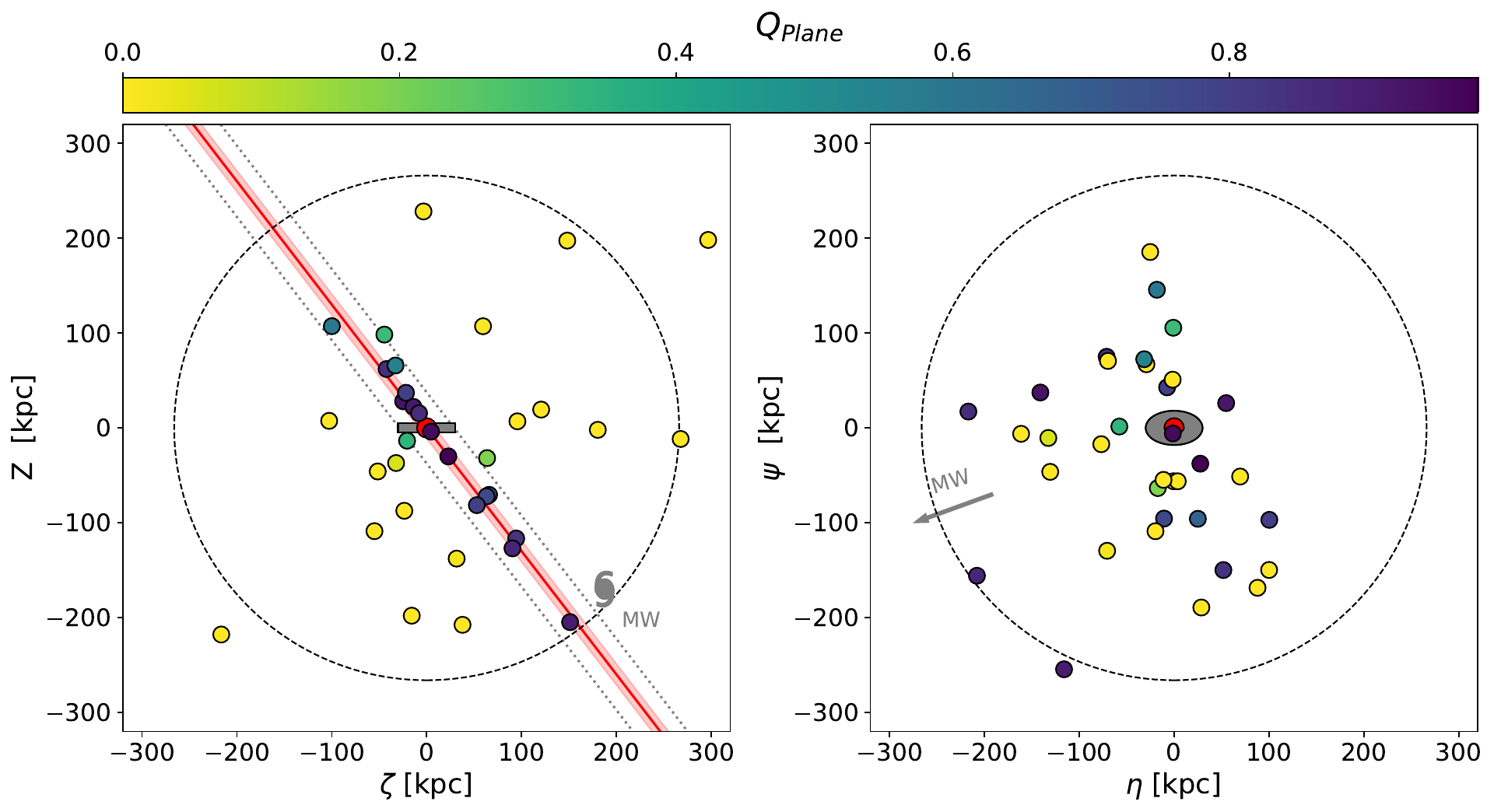}
\caption{Projection of the M31 satellite system along the directions that present an edge-on (left panel) and face-on (right panel) view of the planar structure described in \S~\ref{sec:plane}. The galaxies are color coded by the probability of belonging to the planar component, according to our model. The M31 disc (not to scale) orientation and the position of the MW are also reported. The dashed circle traces the virial radius of M31 \citep[266~kpc,][]{Fardal13,Putman21}. The solid red line, in the left panel, traces the center of our planar model. The red shaded region and the gray dotted lines show the minimum ($\sim 7$kpc) and maximum ($\sim 23$ kpc) thickness we infer in \S~\ref{sec:plane}, respectively.}
\label{fig:plane}
\end{figure*}

Because we use a GMM, we did not make binary decisions on which galaxies belonged to which component. Rather, we assigned each galaxy a probability of belonging to the planar component, defined as:
\begin{equation}
    Q_{Plane} = 2(1-CDF(D_{Plane})),
\end{equation}
 where $CDF(D_{Plane})$ is the cumulative distribution function function, evaluated at the point $D_{Plane}$, of the Gaussian $\mathcal{N}(0,\sqrt{rms^2_{Plane}+\sigma^2_{DP}})$. We used this probability to scale the likelihoods arising from the planar and spherical components, in a similar manner as Eq.~\ref{eq:likelihood}.

We use \texttt{emcee} to explore the parameter space of our model ($rms_{Plane}$, $rms_{Sphere}$), and two parameters for the plane orientation), using uniform priors, with 150 independent walkers randomly distributed in plane orientation. All of them recovered the same predominantly planar component. Viewed from the MW, this is a nearly edge-on feature comprising roughly half of the M31 satellites. Both the orientation and the probable members are in agreement with previous characterizations of the GPoA \citep[e.g., ][]{Conn13,Ibata13,Pawlowski13,Santos-Santos20}.

Given its close to edge-on nature, it is not surprising that the GPoA is preserved with our updated distances. However, with our measurements, we update the structural properties of the plane. We find that the inclination of the plane with respect to the MW is roughly $14^{\circ}$.  This is larger than the original $\sim 1^{\circ}$ value reported in \citet{Ibata13} and more in line with the $12^{\circ}$ figure determined by \citet{Santos-Santos20}.

The left panel of Fig.~\ref{fig:plane} shows a projection of the M31 system along the direction of minimum scatter for this planar structure. Galaxies are color-coded by their probability of being plane members, $Q_{Plane}$ (also provided in Tab.~\ref{tab:DM}). Our model results in a best fit $rms_{Plane} = 23.4^{+11}_{-7.1}$~kpc. This value, however, is somewhat dependent on the details of our model implementation. Given our choice of GMM formalism, and the dependence of $Q_{Plane}$ on $rms_{Plane}$, an accurate measurement of the plane $rms$ thickness depends on the fidelity of the isotropic contamination model. Considering the satellite anisotropy discussed in \S~\ref{sec:Cartesian}, and possible deviations of the radial density profile from the assumed one, a poor fit of the spherically symmetric component could result in an artificial inflation of the best fit $rms_{Plane}$ value, which should be considered an upper limit on the real dispersion of this planar structure. Indeed, if we select only the 15 satellites with $Q_{Plane}>0.5$, we measure a dispersion of only $\sim 7$~kpc around the best fitting plane. This value, on the other hand, should be interpreted as a lower limit, because a selection of the highest-probability members will, by construction, return those galaxies that have the smallest distance to the plane, therefore biasing the measured thickness to lower values. Given these considerations, we find that our limits on the plane thickness are compatible with the values of 12-14~kpc reported in literature \citep{Conn13,Ibata13,Pawlowski13}.

One aspect that should be noted is that, while in this section we have referred to the GPoA as a ``planar" structure, its members do not distribute homogeneously throughout the best-fit plane. This can be appreciated in the right panel of Fig.~\ref{fig:plane}, which shows the view of the M31 system from a direction normal to the GPoA. Here it can be seen that the high-probability GPoA members participate to the general satellite asymmetry described in \S~\ref{sec:Cartesian}. In fact, we find that a similar fraction of galaxies brighter than $M_V=-7.5$ is found on the near side of M31 in the sub-samples with $Q_{Plane}>0.5$ (82\%) and $Q_{Plane}<0.5$ (77\%). Besides the nomenclature implications on whether this structure should be best referred to as an ``arc'', rather than a ``plane'', this degree of asymmetry in the GPoA has relevance for some of its formation models, such as the tidal dwarf scenario \citep[e.g., ][]{Metz07a,Metz07b,Hammer13,Banik22}, which require emergence of this feature at early times and, therefore, should result in an advanced degree of phase-mixing in its constituent members.

While the long-term stability of the GPoA and its significance in the context of $ \rm \Lambda CDM$ have been the subject of discussion for more than a decade, major insights into the nature of this structure will ultimately be enabled by a full orbital analysis of its candidate members. Our program is laying the foundation for this orbital characterization, by providing precise distances now and by establishing an astrometric first epoch for many of the M31 galaxies that have no available proper motion. While proper motions have already been measured for a handful of M31 satellites \citep[e.g., ][]{Sohn20}, members of our team are already in the process of obtaining second epoch imaging for additional galaxies (\hst GO-16273, PI:\ T. Sohn; \textit{JWST} GTO-1305, PI: van der Marel). Only with full phase space information can the plane be fully understood in a cosmological context.

\subsection{The Giant Stellar Stream}
In addition to a virtually complete sample of known M31 satellites, our dataset includes RR Lyrae that are along a line-of-sight to the GSS, which is situated at a projected distance of $\sim 20$~kpc from M31. For this specific field of the GSS, we find $\mu = 24.58\pm0.07$, corresponding to a distance of $D_{M31} = 53.4^{+30}_{-26}$~kpc behind M31 as seen from the MW. Analyzing the same field, \citet{Jeffery11} report a distance modulus of $\mu = 24.52\pm0.19$. 

The GSS \hst\ field  also overlaps a CFHT field of \citet[Field 7]{McConnachie03}, for which they find a distance modulus of $\mu = 24.59\pm0.05$. The \citet{McConnachie03} field is, in turn, contained in the GSS2 field of \citet{Conn16}, for which they report a TRGB distance modulus of $\mu = 24.40^{+0.03}_{-0.02}$.

It is unclear whether the difference with the distance estimated by \citet{Conn16} represents a concern.  The \citet{Conn16} field of view covers a much larger area and averages over a much larger portion of the GSS compared to the GSS \hst\ field. \citet{Conn16} note that this region of the sky is particularly difficult to model, with projected overlap between distinct features and relatively high contamination from the smooth inner halo of M31. Ultimately, this highlights the problem of interpreting the geometric properties of tidal debris even in relatively close galaxies, as low surface brightness and complex morphology can hamper the deprojection of 2D structures to a sufficient level of accuracy. In the case of M31, a more extensive survey of the RR Lyrae populations across the GSS and its neighboring structures could certainly provide precious information for a reliable 3D reconstruction, which in turn will yield deeper insight into the accretion events that produced these tidal features \citep[e.g., ][]{Hammer18}.

\subsection{The NGC~147/NGC~185 Pair}
NGC~147 and NGC~185 are two of the brightest satellite galaxies around M31 and in the Local Group. Due to their close proximity on the sky ($\sim$12~kpc, projected), it has long been suspected that NGC~147 and NGC~185 formed a bound galaxy pair \citep[e.g.,][]{VandenBergh98,Arias16}. Using information from radial velocities, and adopting a separation of $\sim$ 60~kpc, \citet{Geha10} suggested that NGC~147 and NGC~185 are indeed gravitationally bound. More recently, \citet{Sohn20} measured \hst-based proper motions for these galaxies and, through orbital history modeling, were able to rule out NGC~147/185 as a bound galaxy pair. This conclusion was drawn assuming a separation of $\sim$ 89~kpc between the two galaxies. 

Using our distances, we find a separation between NGC~147 and NGC~185 of $83.8^{+25.4}_{-24.6}$~kpc. This is only marginally smaller than previous determinations and the result of \citet{Sohn20} is unaffected. 

\subsection{Candidate Satellites of M33}
As the third largest galaxy in the Local Group, M33 is expected to host its own system of satellite galaxies \citep[e.g.,][]{Dooley17,Bose18,Patel18}. This prediction compares with a paucity of observed dwarf galaxies associated with M33, with the limiting magnitude of existing observations and modest coverage of the M33 virial region likely playing a significant role in the difference \citep{Patel18}. 
The most likely known dwarf galaxy associated with M33 is And~{\sc XXII} as it has a close projected separation ($\sim43$~kpc). However, the physical association between M33 and And~{\sc XXII} ultimately depends on their deprojected distance. Using CFHT photometry, \citet{Chapman13}, estimated a separation of $59^{+21}_{-14}$~kpc between And~{\sc XXII} and M33, and suggested that the two systems are currently gravitationally bound.

Using our distances, we find a separation of $113.4^{+31.2}_{-29.7}$~kpc between these two galaxies. While this is still within the estimated virial radius of M33 \citep[161~kpc, ][]{Patel17}, this new measurement makes the status of And~{\sc XXII} as a bound satellite of M33 ultimately much more sensitive to the orbital history of the latter. Orbital solutions that exclude recent close encounters with M31 \citep[e.g., ][]{Patel17,VanderMarel19}, would be compatible with And~{\sc XXII} being a current satellite of M33. However, models such as those put forward by \citet{McConnachie09} and \citet{Putman09} advocate for a recent $<100$~kpc pericenter passage of M33 around M31.  They imply that And~{\sc XXII}, if it ever was associated with M33, is now likely unbound from its former host. Conversely, the assumption that And~{\sc XXII} is currently associated with M33 would disfavor the hypothesis of a close M31-M33 interaction. Ultimately, the status of And~{\sc XXII}, and its relevance in the context of M33's orbital history, will become much clearer when measurements of the proper motion for this dwarf galaxy become available.

With regards to the orbital history of M33, it should be noted that, while our updated distance to M31 is consistent with the value of $770\pm40$ kpc used in recent models of the M31-M33 interaction \citep[e.g., ][]{Patel17,VanderMarel19}, the distance to M33 has changed more significantly from the value of $794\pm23$ kpc used in those models. The resulting M31-M33 distance is therefore increased by $\sim25$ kpc. Such change is unlikely to have profound consequences on the orbit of M33 but work is already undergoing to include these new distance determinations into an updated dynamical modeling of the M31-M33 pair (Patel et al., in prep.).

Recently, a new tentative satellite of M33 was identified with the discovery of Pisces~VII \citep{Martinez-Delgado22}. With an estimated TRGB distance of $1.0^{+0.3}_{-0.2}$~Mpc, it is currently not clear whether Pisces~VII is an isolated galaxy \citep[like And~XVIII,][]{Makarova17} or is indeed associated with the M31/M33 system. Unfortunately, due to its very recent discovery, this galaxy is not present in our sample. The low luminosity of Pisces~VII ($M_V=-6.8\pm0.2$), especially in light of the results of \S~\ref{sec:TRGB}, prevents a more robust localization on the basis of the TRGB alone. The only way to measure a secure distance to this galaxy is through cadenced observations like those presented in this paper.

\section{Conclusions}

We presented homogeneous distances to 38 satellite systems that orbit M31 and M31 itself, using $>700$ \hst\ orbits analyzed as part of the \hst\ Survey of M31 Satellite Galaxies (GO-15902; PI:\ D. Weisz).  This effort is the largest homogeneous set of RR Lyrae-based distances for nearby galaxies. We summarize the main takeaways of this paper as follows:
\newline
\newline
i) We identified and characterized $\>4700$ RR Lyrae stars in 39 stellar systems, including M31, M33, and almost all known dwarf galaxies within the virial radius of M31. We measure pulsation periods to a typical precision of 0.01~days, mean magnitudes to 0.04~mag, and pulsation amplitudes to 0.11~mag. We unveil a diversity of RR Lyrae demographics in our sample, that can only partially be explained by luminosity effects and observational completeness. This suggests a variety of formation/enrichment histories, which we will quantify in subsequent papers.
\newline
\newline
ii) From the RR Lyrae measurements, we derived homogeneous distances to all the stellar systems in our sample. We used dust-independent PWZ calibrations, anchored to the \textit{Gaia} eDR3 distance scale, which enabled us to measure distances relative to the MW to a precision of $\sim20$~kpc ($\sim3$\%) for all the galaxies in our sample, including those as faint as $M_V = -6$. We computed relative distances with respect to M31, which we determined to a typical precision of $\sim 10$~kpc (8\%). Through extensive tests we quantified our typical systematic uncertainties to be $\sim 3.5\%$, i.e., $\sim 25$~kpc on the distances from the MW and $\sim 4$~kpc on the relative distances to M31.  With our new distances, we provide updated satellite galaxy luminosities and sizes.
\newline
\newline
iii) We found that our RR Lyrae-based distances and the TRGB distances of bright galaxies in our sample agree within 0.05-0.07 mag. For stellar systems with $M_V\gtrsim -9.5$, we report that the TRGB systematically over-predicts distances compared to the RR Lyrae, quite substantially in some cases.  This indicates that stochastic sampling of the TRGB is a serious problem below this luminosity. This highlights the importance of acquiring variable star-based distances for the many faint nearby galaxies expected to be discovered by upcoming deep photometric surveys (e.g., through the Rubin and Roman observatories).
\newline
\newline
iv) We mapped the 3D structure of the M31 satellite system and characterized known substructures in the distribution of dwarfs around M31. We confirmed the existence of a stark asymmetry in the satellite distribution, with more than 80\% of the satellites being located in the direction of the MW. We affirm the existence of a 7-23 kpc thick planar structure that comprises roughly half of our sample. This structure shares the overall asymmetry of the satellite distribution and mostly populates one side of the M31-centric plane it aligns with.
\newline
\newline
v) We used our updated distances to examine the physical association of close galaxy pairs. We derived a deprojected distance of $\sim 84$~kpc between NGC~147 and NGC~185, supporting the claim that this galaxy pair is currently not gravitationally bound. Similarly, we measured the separation between M33 and And~{\sc XXII} to be $\sim 113$~kpc, placing the small dwarf in the outskirts of M33's virial region.  We  discuss the status of And~{\sc XXII} as a bound satellite of M33 in light of putative past M33-M31 interactions. The proper motions of And~{\sc XXII} will be especially informative for the orbital history of M33.
\newline
\newline
The precise and homogeneous analysis of RR Lyrae, and the set of uniform distances they enable, are at the foundation of our M31 survey.   They are central to star formation history and orbital history science, and also enable a better understanding of the relationship between variable stars and galaxy stellar populations.  This work also represents a significant step in bringing our knowledge of the M31 satellite system onto comparable footing to that of faint galaxies in the MW halo.

\begin{acknowledgments}
AS wishes to thank Y. Zheng for useful discussions at various stages of this work. Support for this work was provided by NASA through grants GO-13768, GO-15746, GO-15902, AR-16159, and GO-16273 from the Space Telescope Science Institute, which is operated
by AURA, Inc., under NASA contract NAS5-26555. MCC acknowledges support though NSF grant AST-1815475. MBK acknowledges support from NSF CAREER award AST-1752913, NSF grants AST-1910346 and AST-2108962, NASA grant NNX17AG29G, and HST-AR-15006, HST-AR-15809, HST-GO-15658, HST-GO-15901, HST-GO-15902, HST-AR-16159, and HST-GO-16226 from STScI. This research has made use of NASA’s Astrophysics Data System Bibliographic Services. All the \hst\ data used in this paper can be found in MAST: \dataset[10.17909/jb41-ex86]{http://dx.doi.org/10.17909/jb41-ex86}

\end{acknowledgments}

%

\vspace{5mm}
\facilities{ \hst\ (ACS) }


\software{ This research made use of routines and modules from the following software packages: \texttt{Astropy} \citep{Astropy}, \texttt{DOLPHOT} \citep{Dolphin16}, \texttt{emcee} \citep{Foreman-Mackey13}, \texttt{IPython} \citep{IPython}, \texttt{Matplotlib} \citep{Matplotlib}, \texttt{NumPy} \citep{Numpy}, \texttt{Pandas} \citep{Pandas}, \texttt{Psearch} \citep{Saha17} and \texttt{SciPy} \citep{Scipy}
          }



\newpage
\bibliographystyle{aasjournal}
\bibliography{Bibliography.bib}{}



\end{document}